\documentclass[pra,showpacs,twocolumn,superscriptaddress]{revtex4}

\usepackage{graphicx}
\usepackage{multirow}
\usepackage[usenames]{color}
\usepackage{nicefrac}
\usepackage{amsmath}
\usepackage{bm}

\begin{document}

\title{Tools for designing atom interferometers in a microgravity environment}

\author{Elizabeth Ashwood}
\affiliation{Department of Physics, Georgia Southern University,
Statesboro, GA 30460--8031 USA}

\author{Ed Wesley Wells} 
\affiliation{Department of Physics, Georgia Southern University,
Statesboro, GA 30460--8031 USA}

\author{Doga Murat Kurkcuoglu} 
\affiliation{Department of Physics, Georgia Southern University,
Statesboro, GA 30460--8031 USA}

\author{Robert Colson Sapp} 
\affiliation{Department of Physics, Georgia Southern University,
Statesboro, GA 30460--8031 USA}

\author{Charles W.\ Clark}
\affiliation{Joint Quantum Institute, National Institute of Standards 
and Technology and the University of Maryland, Gaithersburg, MD 20899, USA}

\author{Mark Edwards}
\affiliation{Department of Physics, Georgia Southern University,
Statesboro, GA 30460--8031 USA}
\affiliation{Joint Quantum Institute, National Institute of Standards 
and Technology and the University of Maryland, Gaithersburg, MD 20899, USA}

\date{\today}

\begin{abstract}
We present a variational model suitable for rapid {\em preliminary design} of atom interferometers in a microgravity environment.  The model approximates the solution of the 3D rotating--frame Gross--Pitaevskii equation (GPE) as the sum of $N_{c}$ Gaussian clouds. Each Gaussian cloud is assumed to have time--dependent center positions, widths, and linear and quadratic phase parameters.  We applied the Lagrangian Variational Method (LVM) with this trial wave function to derive equations of motion for these parameters that can be adapted {\em to any external potential}.  We also present a 1D version of this variational model. As an example we apply the model to a 1D atom interferometry scheme for measuring Newton's gravitational constant, $G$, in a microgravity environment.  We show how the LVM model can (1) constrain the experimental parameter space size, (2) show how the value of $G$ can be obtained from the experimental conditions and interference pattern characteristics, and (3) show how to improve the sensitivity of the measurement and construct a preliminary error budget.
\end{abstract}


\maketitle

\section{Introduction}
\label{intro}

Atom interferometers (AIs) have been used in many applications\,\cite{ai_book,RevModPhys.81.1051}. In particular this is the case for gravitational measurements, such as the determination of Earth's gravitational acceleration,\,$g$,\,\cite{peters} and its gradients\,\cite{PhysRevA.65.033608,1367-2630-12-9-095009}. Atom interferometers have also been proposed for use in Sagnac gyroscopes\,\cite{ClassQuantumGrav.17.2385,PhysRevLett.97.010402}, in testing general relativity\,\cite{PhysRevLett.93.240404,equivalence,PhysRevLett.113.023005}, in searching for dark energy\,\cite{dark_energy}, and measuring gravity waves\,\cite{PhysRevD.78.122002,PhysRevLett.110.171102}.  The Newtonian constant of gravitation, $G$, has also been measured with AIs\,\cite{big_G,PhysRevLett.100.050801}.  

In a typical atom interferometer involving a gaseous Bose--Einstein condensate (BEC), the basic sequence consists of splitting the condensate into two or more clouds that then experience different environments while separated.  Some differing property of these environments generates a phase difference between the clouds while they are separated. The phase difference produces an interference pattern when the clouds are reunited and this pattern can be used to infer the value of this property.

An emerging area of AI applications is to systems that are freely falling in a gravitational field. Such systems offer the prospect of much longer interrogation times than are possible in fixed terrestrial laboratories, where atoms are subject to constant vertical acceleration.  Such ``microgravity'' environments have been created in drop towers~\cite{drop_towers}, sounding rockets~\cite{sounding_rocket_ai}, and recently aboard the International Space Station (ISS). 

The National Aeronautics and Space Administration (NASA) has deployed its Cold Atom Laboratory (CAL)~\cite{nasa_cal} to the ISS enabling future atom--interferometry experiments to be performed on Bose--Einstein condensates there.  This will provide a user facility in which atom--interferometry experiments can be conducted in a microgravity environment.

The basic AI sequence of splitting, reuniting, and splitting the BEC again to produce an interference pattern can be viewed as a sequence of sudden splits of the condensate alternating with periods where the condensate clouds evolve freely according to their local environments. In general, the performance of AIs of this type depends on several factors including the number of interferometer pathways, the area of the interferometer, and the interrogation time (the time the clouds spend in the different environments).  

AI performance can be enhanced both by increasing any or all of these quantities.  Specific performance enhancements can also be achieved by designing different AI sequences as was done recently with neutron interferometers\,\cite{PhysRevA.79.053635}.  Differential AI experiments, where the same AI sequence is carried out on identical condensates at the same time, can also be used in making precision measurements.

Enhancing AI performance by increasing interrogation time is especially possible in a microgravity environment.  The extended interrogation times available in microgravity are also valuable for precision measurement applications.  Microgravity environments in general and the CAL in particular are important platforms for making precision measurements.  To further this end, designing new AI sequences that take advantage of microgravity would be highly desirable.

Designing new AI sequences involving BECs suitable for precision measurements conducted in microgravity environments presents a demanding numerical challenge.  Detailed modeling of condensate behavior for AI sequences having long interrogation times and/or multiple pathways requires large amounts of computer time and storage assuming that the condensate obeys the Gross-Pitaevskii equation (GPE). Many different possible sequences would also need to be simulated and the results evaluated.  Thus a method for {\em rapid simulation} of a given AI sequence would be an extremely useful tool for AI design.  Such a tool would also be helpful in determining how the quantity measured by the AI could be extracted from the resulting interference patterns.

This paper presents a tool suitable for preliminary design of candidate AI sequences.  The tool can provide rapid approximate modeling of condensate behavior between the splits of an AI sequence where this evolution is assumed to be governed by the GPE.  This model is based on the Lagrangian Variational Method (LVM) where the assumed trial wave function approximates the GPE solution as a set of $N_{c}$ separate but possibly overlapping Gaussian clouds.  We emphasize that the final choice of an AI sequence would have to be validated by simulating it with the GPE.

In section \ref{LVM} we describe the general Lagrangian Variational Method, introduce a set of scaled units suitable for computation, and present the trial wave functions for both the one-dimensional and three-dimensional versions of the method.  In section \ref{eoms} we derive the equations of motion for the variational parameters and show that these can be cast in terms of the space and width gradients of a ``variational potential'' which is just the expectation value of the external potential over the trial wave function.  

In section \ref{bigG} we describe an example of an AI scheme that could be used to measure $G$ in microgravity. In section \ref{lvm_analysis} we apply the LVM model to this AI scheme. We compare a GPE simulation of this scheme with the LVM simulation.  We also use the model to obtain an expression for the LVM interference pattern, show how $G$ can be extracted from this pattern, and how the model can facilitate preliminary design of an AI sequence.  Finally, we present a summary of the work in section \ref{summary}.

\section{The Lagrangian Variational Method and Trial Wave Function}
\label{LVM}

\subsection{The general Lagrangian variational method}
\label{lvm_general}

The Lagrangian Variational Method provides approximations to the solutions of the time--dependent Gross--Pitaevskii equation\,\cite{PhysRevA.56.1424, PhysRevE.86.056710,0953-4075-39-6-L02,PhysRevA.65.043614,PhysRevA.79.053620}.  In three dimensions and in the rotating frame, this equation has the form\,\cite{Pitaevskii_and_Stringari}:
\begin{eqnarray}
i\hbar\frac{\partial\Phi}{\partial t}
&=&
-\frac{\hbar^{2}}{2M}\nabla^{2}\Phi + 
V_{\rm ext}({\bf r},t)\Phi +
g_{3D}N\left|\Phi\right|^{2}\Phi\nonumber\\
&+&
i\hbar{\bf \Omega}\cdot\left({\bf r}\times{\bf\nabla}\right)\Phi,
\end{eqnarray}
where $\Phi({\bf r},t)$ is the condensate wave function, $M$ is the mass of a condensate atom, $N$ is the number of atoms in the condensate, $g_{3D}=4\pi\hbar^{2}a_{s}/M$ measures the strength of the atom--atom scattering with $a_{s}$ being the scattering length, $V_{\rm ext}({\bf r},t)$ is the potential exerted on a condensate atom by external fields, and ${\bf\Omega}$ is the angular velocity of the rotating frame.

The LVM is based on the fact that the GPE can be derived as the Euler--Lagrange equation of motion produced by the following Lagrangian density:
\begin{eqnarray}
{\cal L}[\Phi^{\ast}]
&=& 
\hbar
{\rm Im}\left\{\Phi^{\ast}
\Phi_{t}\right\} +
\frac{\hbar^{2}}{2M}
{\bf\nabla}\Phi^{\ast}\cdot{\bf\nabla}\Phi\nonumber\\ 
&+&
V_{\rm ext}({\bf r},t)\Phi^{\ast}\Phi + 
\tfrac{1}{2}g_{3D}N(\Phi)^{2}(\Phi^{\ast})^{2}\nonumber\\ 
&+&
\Phi
\left(
{\bf\Omega}\cdot\hat{\bf L}
\right)\Phi^{\ast},
\end{eqnarray}
where $\hat{\bf L}$ is the single--particle angular momentum operator.

This Lagrangian density along with the following Euler--Lagrange equation of motion produces the GPE:
\begin{equation}
\sum_{\eta=x,y,z,t}
\frac{\partial}{\partial\eta}
\left(\frac{\partial{\cal L}}{\partial\Phi^{\ast}_{\eta}}\right) -
\frac{\partial{\cal L}}{\partial\Phi^{\ast}}  = 0,
\quad{\rm where}\quad
\Phi_{\eta} \equiv \frac{\partial\Phi}{\partial\eta}.
\end{equation}

The Lagrangian Variational Method consists of devising a {\em trial} wave function,
\begin{equation}
\Phi^{\rm trial}({\bf r},t) = \Phi^{\rm trial}(q_{1}(t),\dots,q_{n}(t);{\bf r})
\end{equation}
where the $\{q_{i}(t)\}$, $i= 1,\dots,n$ are {\em variational parameters} that only depend on the time, $t$.  The equations of motion of these variational parameters are derived by computing the ordinary Lagrangian:
\begin{equation}
L(q_{1}(t),\dots,q_{n}(t)) = 
\int\,d^{3}r
{\cal L}[\left(\Phi^{\rm trial}\right)^{\ast}]
\end{equation}
and then using the standard Euler--Lagrange equations,
\begin{equation}
\frac{d}{dt}
\left(\frac{\partial L}{\partial \dot{q}_{k}}\right) - 
\frac{\partial L}{\partial q_{k}} = 0,
\quad
k = 1,\dots,n
\label{standard_EL}
\end{equation}
to produce an equation of motion associated with each variational parameter.

\subsection{Scaled units}
\label{scaled_units}

We can simplify the equations produced by the above method by introducing a set of units appropriate to the problem and a set of scaled variables (both independent and dependent). The scaled variables are defined by first choosing a {\em length unit}, $L_{0}$, and then defining energy and time units in terms of $L_{0}$:
\begin{equation}
E_{0} \equiv \frac{\hbar^{2}}{2ML_{0}^{2}}
\quad{\rm and}\quad
T_{0} \equiv \frac{\hbar}{E_{0}} = \frac{2ML_{0}^{2}}{\hbar}.
\end{equation}
We then introduce scaled variables which are generally denoted by barred quantities. These consist of scaled space and time coordinates:
\begin{equation}
\bar{x} \equiv \frac{x}{L_{0}}\quad
\bar{y} \equiv \frac{y}{L_{0}}\quad
\bar{z} \equiv \frac{z}{L_{0}}\quad{\rm and}\quad
\bar{t} \equiv \frac{t}{T_{0}}.
\end{equation}
We also introduce the scaled condensate wave function for the solution of the
GPE:
\begin{equation}
\Phi({\bf r},t) = L_{0}^{-3/2}\Psi({\bf \bar{r}},\bar{t}).
\end{equation}
We can express the original GPE in terms of scaled quantities and this can be done for the Lagrangian density and its associated Euler--Lagrange equation as well.

In terms of scaled quantities, the rotating--frame GPE becomes:
\begin{eqnarray}
i\frac{\partial\Psi}{\partial\bar{t}} 
&=& 
-\bar{\nabla}^{2}\Psi + 
 \bar{V}_{\rm ext}({\bf\bar{r}},\bar{t})\Psi +
 \bar{g}_{3D}N\left|\Psi\right|^{2}\Psi\nonumber\\ 
&+&
i\bar{\Omega}_{z}
 \left(
 \bar{x}\frac{\partial\Psi}{\partial\bar{y}} -
 \bar{y}\frac{\partial\Psi}{\partial\bar{x}}
 \right)
\label{scaled_3d_GPE}
\end{eqnarray}
where we are considering the form of the rotating--frame GPE for the special case of rotation around the $z$ axis.  In the above we have $g_{3D}\equiv\bar{g}_{3D}E_{0}L_{0}^{3}$, $V_{\rm ext}({\bf\bar{r}},\bar{t})=\bar{V}_{\rm ext}({\bf r},t)E_{0}$ and $\Omega_{z}=\bar{\Omega}_{z}/T_{0}$.  The scaled Lagrangian density for this version of the GPE becomes
\begin{eqnarray}
\bar{\cal L}^{(3D)}
\left[\Psi^{\ast}\right] 
&=& 
{\rm Im}\left\{\Psi^{\ast}\Psi_{\bar{t}}\right\} +
\bar{\nabla}\Psi^{\ast}\cdot\bar{\nabla}\Psi + 
\bar{V}_{\rm ext}({\bf\bar{r}},\bar{t})\Psi\Psi^{\ast}\nonumber\\
&+&
\tfrac{1}{2}\bar{g}N\left(\Psi^{\ast}\right)^{2}\left(\Psi\right)^{2} +
i\bar{\Omega}_{z}\Psi
\left(\bar{y}\Psi_{x}^{\ast} - \bar{x}\Psi_{y}^{\ast}\right)\nonumber\\
&\equiv&
\bar{\cal L}^{(3D)}_{1}\left[\Psi^{\ast}\right] +
\bar{\cal L}^{(3D)}_{2}\left[\Psi^{\ast}\right] +
\bar{\cal L}^{(3D)}_{3}\left[\Psi^{\ast}\right]\nonumber\\ 
&+&
\bar{\cal L}^{(3D)}_{4}\left[\Psi^{\ast}\right] +
\bar{\cal L}^{(3D)}_{5}\left[\Psi^{\ast}\right]
\label{scaled_3d_Lag}
\end{eqnarray}
and the scaled Euler--Lagrange equation is given by
\begin{equation}
\sum_{\eta=x,y,z,t}
\frac{\partial}{\partial\bar{\eta}}
\left(
\frac{\partial\bar{\cal L}^{(3D)}}{\partial\Psi^{\ast}_{\bar{\eta}}}
\right) -
\frac{\partial\bar{\cal L}^{(3D)}}{\partial\Psi^{\ast}} = 0.
\label{EL_3D}
\end{equation}
It is straightforward to insert Eq.\ (\ref{scaled_3d_Lag}) into Eq.\ (\ref{EL_3D}) and obtain Eq.\ (\ref{scaled_3d_GPE}).  The 3D Lagrangian is computed by integrating over all 3D space:
\begin{eqnarray}
\bar{L}^{(3D)}
\left[\bf{x},\bf{w},\bm{\alpha},\bm{\beta}\right]
&=& 
\int\,d^{3}\bar{r}
\bar{\cal L}^{(3D)}\left[\Psi^{\ast}\right]\nonumber\\
&=& 
\sum_{k=1}^{5}\int\,d^{3}\bar{r}\bar{\cal L}^{(3D)}_{k}\nonumber\\
&\equiv&
\bar{L}^{(3D)}_{1} +
\bar{L}^{(3D)}_{2} +
\bar{L}^{(3D)}_{3}\nonumber\\ 
&+&
\bar{L}^{(3D)}_{4} +
\bar{L}^{(3D)}_{5}.
\label{3d_lag_int}
\end{eqnarray}
In the above have denoted the dependence of $\bar{L}$ on the different types of variational parameters that will appear in the trial wave function defined in the next section.  These are: the cloud center coordinates by $\bm{x}\equiv \left(\bar{x}_{1},\dots,\bar{z}_{N_{c}}\right)$, the cloud widths denoted by $\bm{w}\equiv \left(\bar{w}_{1x},\dots,\bar{w}_{N_{c}z}\right)$, the linear phase coefficients by $\bm{\alpha}\equiv \left(\alpha_{1x},\dots,\alpha_{N_{c}z}\right)$ and the quadratic phase coefficients by $\bm{\beta}\equiv \left(\beta_{1x},\dots,\beta_{N_{c}z}\right)$.

The LVM formulation is straightforward to apply in one--dimension where we denote the 1D GPE solution by $\phi(x,t)$.  For scaled units we will write $\phi(x,t)\equiv L_{0}^{-1/2}\psi(\bar{x},\bar{t})$.  In terms of scaled quantities, the 1D GPE becomes:
\begin{equation}
i\frac{\partial\psi}{\partial\bar{t}} = 
-\frac{\partial^{2}\psi}{\partial\bar{x}^{2}} + 
 \bar{V}_{\rm ext}(\bar{x},\bar{t})\psi +
 \bar{g}_{1D}N\left|\psi\right|^{2}\psi.
\end{equation}
where $g_{1D} \equiv \bar{g}_{1D}E_{0}L_{0}$ and $V_{\rm ext}(\bar{x},\bar{t})=
\bar{V}_{\rm ext}(x,t)/E_{0}$.  The 1D scaled Lagrangian density becomes
\begin{eqnarray}
\bar{\cal L}^{(1D)}
\left[\psi^{\ast}\right] 
&=& 
{\rm Im}
\left\{\psi^{\ast}\psi_{\bar{t}}\right\} +
\psi^{\ast}_{\bar{x}}\psi + 
\bar{V}_{\rm ext}(\bar{x},\bar{t})\psi\psi^{\ast}\nonumber\\
&+&
\tfrac{1}{2}\bar{g}_{1D}N
\left(\psi^{\ast}\right)^{2}\left(\psi\right)^{2}
\end{eqnarray}
and the 1D scaled Euler--Lagrange equation is given by
\begin{equation}
\frac{\partial}{\partial\bar{x}}
\left(
\frac{\partial\bar{\cal L}^{(1D)}}{\partial\psi^{\ast}_{\bar{x}}}
\right) +
\frac{\partial}{\partial\bar{t}}
\left(
\frac{\partial\bar{\cal L}^{(1D)}}{\partial\psi^{\ast}_{\bar{t}}}
\right) -
\frac{\partial\bar{\cal L}^{(1D)}}{\partial\psi^{\ast}} = 0.
\end{equation}
We can calculate the Lagrangian associated with this trial wave function
by integrating $\bar{\cal L}^{(1D)}$ over all 1D space:
\begin{equation}
\bar{L}^{(1D)}\left[\bf{x},\bf{w},\bm{\alpha},\bm{\beta}\right] = 
\int_{-\infty}^{\infty}d\bar{x}\,
\bar{\cal L}^{(1D)}
\left[
\psi^{\ast}
\right].
\label{1d_lag_int}
\end{equation}
This Lagrangian will depend only on the variational parameters contained in the trial wave function. The 1D equations of motion are found with the standard Euler--Lagrange equations.  Next we will present the 1D and 3D LVM trial wave functions.

\subsection{$N_{c}$ 3D LVM trial wave function}
\label{3d_LVM_trial_wf}

In the 3D, $N_{c}$--gaussian model we take the trial wave function to be a sum of $N_{c}$ three--dimensional Gaussian clouds.  The $j^{th}$ cloud has its own initial momentum, ${\bf\bar{k}}_{j}$ and set of variational parameters. These parameters are the cartesian coordinates of the cloud center: $\bar{x}_{j}$, $\bar{y}_{j}$, and $\bar{z}_{j}$; the widths along the $x$, $y$, and $z$ directions: $\bar{w}_{jx}$, $\bar{w}_{jy}$, and $\bar{w}_{jz}$; the linear phase coefficients along the $x$, $y$, and $z$ directions: $\bar{\alpha}_{jx}$, $\bar{\alpha}_{jy}$, and $\bar{\alpha}_{jz}$; and the quadratic phase coefficients along the $x$, $y$, and $z$ directions: $\bar{\beta}_{jx}$, $\bar{\beta}_{jy}$, and $\bar{\beta}_{jz}$.  The $j^{th}$ cloud also has its own normalization coefficient, $A_{j}$, which will be eliminated by fixing the number of atoms in each cloud.

The mathematical form (in scaled units) of the trial wave function is the following:
\begin{equation}
\Psi({\bf \bar{r}},\bar{t}) = 
\frac{1}{\sqrt{N_{c}}}\sum_{j=1}^{N_{c}}
A_{j}(\bar{t})
e^{f_{j}({\bf\bar{r}},\bar{t}) + i{\bf\bar{k}}_{j}\cdot{\bf\bar{r}}}
\label{trial_wf_3d}
\end{equation}
where
\begin{eqnarray}
f_{j}({\bf\bar{r}},\bar{t}) 
&=&
\sum_{\eta = x,y,z}
\left(
-\frac{(\bar{\eta}-\bar{\eta}_{j})^{2}}{2\bar{w}_{j\eta}^{2}} +
i\bar{\alpha}_{j\eta}\bar{\eta} + 
i\bar{\beta}_{j\eta}\bar{\eta}^{2}
\right).\nonumber\\
\label{trial_exp_3d}
\end{eqnarray}
This trial wave function will be used below in deriving the 3D Lagrangian function from which the 3D equations of motion will be obtained.

\subsection{$N_{c}$ 1D LVM trial wave function}
\label{1d_LVM_trial_wf}

In the 1D, $N_{c}$--gaussian--cloud model we take the trial wave function to be a sum of $N_{c}$ one--dimensional Gaussian clouds.  The $j^{th}$ cloud has its own initial momentum, $\bar{k}_{j}$ and set of variational parameters. These parameters are the cartesian coordinate of the cloud center, $\bar{x}_{j}$, the cloud width, $\bar{w}_{j}$, the linear phase coefficient, $\bar{\alpha}_{j}$, and the quadratic phase coefficient, $\bar{\beta}_{j}$. The $j^{th}$ cloud also has its own normalization coefficient, $A_{j}$, which will be eliminated by fixing the number of atoms in each cloud.

The mathematical form (in scaled units) of the trial wave function is the following:
\begin{equation}
\psi(\bar{x},\bar{t}) = 
\frac{1}{\sqrt{N_{c}}}\sum_{j=1}^{N_{c}}
A_{j}(\bar{t})
e^{f_{j}(\bar{x},\bar{t}) + i\bar{k}_{j}\bar{x}}
\label{1d_trial_wf}
\end{equation}
where
\begin{equation}
f_{j}(\bar{x},\bar{t}) =
-\frac{(\bar{x}-\bar{x}_{j}(\bar{t}))^{2}}{2\bar{w}_{j}^{2}(\bar{t})} +
i\bar{\alpha}_{j}(\bar{t})\bar{x} +
i\bar{\beta}_{j}(\bar{t})\bar{x}^{2}
\label{1d_trial_exp}
\end{equation}

\section{The LVM Equations of Motion}
\label{eoms}

In this section we derive the 3D and 1D LVM equations of motion.  We first describe our assumed constraints placed on the trial wave function, then we sketch the derivation of the 3D and 1D Lagrangians for the given trial wave functions and finally sketch the derivation of the final LVM equations of motion. Most of the details are relegated to appendices. The terms in these Lagrangians containing the external potential $\bar{V}_{\rm ext}$ only depend on the cloud centers, ${\bf x}$, and widths, ${\bf w}$, no matter the form of the potential. This is due solely to our choice of trial wave function.  Thus the 1D and 3D Lagrangians can be written in terms of a ``variational'' external potential, $U_{\rm ext}({\bf x},{\bf w})$. This enables the terms in the final equations of motion that account for the external potential to be written in terms of the ${\bf x}$ and ${\bf w}$ gradients of this potential.  Hence the final equations of motion can be written in a form independent of the particular $V_{\rm ext}$. 

\subsection{Constraints on the wave function}
\label{wf_constraints}

Here we make several assumptions about the physical system which have material effects on the values of the variational parameters.  These are as follows:
\begin{enumerate}
  \item We assume that each of the $N_{c}$ clouds are moving at sufficiently different velocities such that any integral of a quantity containing a factor like $e^{i\left(\bar{k}_{j\eta}-\bar{k}_{j^{\prime}\eta}\right)\bar{\eta}}$ where $j\ne j^{\prime}$ can be neglected.  If the clouds move with sufficiently different velocities, these factors will be rapidly oscillating and their integrals can be neglected.
  \item The amplitudes, $A_{j}(\bar{t})$, are all real.  Here we assume that the phase of each cloud's wave function is quadratic so that there is no phase is subsumed by the amplitudes. The quadratic phase enables the cloud centers to move and the cloud widths to expand and contract.
  \item The number of atoms in each cloud is fixed.  Clouds do not exchange atoms.  This plus the normalization condition, fixes a relationship (derived below) between $A_{j}$ and the widths $\bar{w}_{j\eta}$ where $\eta = x,y,z$.
\end{enumerate}
We can use these assumptions plus the normalization condition on the trial wave function to derive conditions that constrain the values of the $A_{j}$.

To find these conditions we require that the full trial wave function be normalized to unity:
\begin{eqnarray}
1 
&=&
\int\,d^{3}\bar{r}\left|\Psi({\bf\bar{r}},\bar{t})\right|^{2}\nonumber\\ 
&=&
\frac{1}{N_{c}}
\sum_{j=1}^{N_{c}}
A_{j}^{2}(\bar{t})
\left(\pi^{3/2}
\bar{w}_{jx}(\bar{t})
\bar{w}_{jy}(\bar{t})
\bar{w}_{jz}(\bar{t})
\right),
\label{full_norm}
\end{eqnarray}
where we have used assumptions 1 and 2 above to reduce the normalization condition to a sum of Gaussian integrals which are easily evaluated.  This last expression is the condition for the trial wave function to be normalized.  However, our assumption that the number of atoms in each cloud is fixed adds a further restriction to the above expression.  That is that each cloud is individually normalized.  This gives, finally,
\begin{equation}
A_{j}^{2}(\bar{t})
\pi^{3/2}
\bar{w}_{jx}(\bar{t})
\bar{w}_{jy}(\bar{t})
\bar{w}_{jz}(\bar{t}) = 1,
\quad
j = 1,\dots,N_{c}.
\label{3d_cloud_norm}
\end{equation}
These constraints together automatically satisfy Eq.\ (\ref{full_norm}) and enable the elimination of all of the $A_{j}$ in the final 3D Lagrangian.  A similar condition on the norm of each cloud in the 1D case can be derived and the result is
\begin{equation}
A_{j}^{2}(\bar{t})
\pi^{1/2}
\bar{w}_{j}(\bar{t}) = 1,
\quad
j = 1,\dots,N_{c}.
\label{1d_cloud_norm}
\end{equation}
Using these conditions we can now derive the 3D and 1D Lagrangians.  

\subsection{The LVM model Lagrangians}
\label{3d_1d_lags}

The LVM Lagrangians for the particular choice of trial wave function given above can be derived by inserting Eq.\,(\ref{trial_wf_3d}) into Eq.\,(\ref{3d_lag_int}) and performing the required integrations.  The details are given in Appendix\,\ref{3d_LVM_Lagrangian}. The result for the 3D Lagrangian is
\begin{eqnarray}
\bar{L}^{(3D)} 
&=&
\frac{1}{N_{c}}\sum_{j=1}^{N_{c}}
\sum_{\eta=x,y,z}
\Bigg[\dot{\bar{\alpha}}_{j\eta^{\prime}}\bar{\eta}_{j} + 
\dot{\bar{\beta}}_{j\eta}
\left(
\bar{\eta}^{2}_{j} +
\frac{1}{2}\bar{w}_{j\eta}^{2}
\right)\nonumber\\
&+&
\frac{1}{2\bar{w}_{j\eta}^{2}} + 
2\bar{\beta}_{j\eta}^{2}
\bar{w}_{j\eta}^{2} +
\Big(
2\bar{\beta}_{j\eta}\bar{\eta}_{j} +
\bar{\alpha}_{j\eta} +
\bar{k}_{j\eta}
\Big)^{2}
\Bigg]\nonumber\\ 
&+&
\frac{\bar{\Omega}_{z}}{N_{c}}
\sum_{j=1}^{N_{c}}
\Bigg[
\bar{y}_{j}
\left(
\bar{\alpha}_{jx} + 
\bar{k}_{jx} +
2\bar{\beta}_{jx}\bar{x}_{j}
\right) -
\bar{x}_{j}
\big(
\bar{\alpha}_{jy}\nonumber\\ 
&+& 
\bar{k}_{jy} +
2\bar{\beta}_{jy}\bar{y}_{j}
\big)
\Bigg] +
\frac{1}{2N_{c}}
\Big(
\bar{U}^{(3D)}_{\rm ext}({\bf x},{\bf w})\nonumber\\
&+& 
\bar{U}^{(3D)}_{\rm int}({\bf x},{\bf w})
\Big).
\label{L_final_3d}
\end{eqnarray}
The variational potentials are defined as
\begin{eqnarray}
\bar{U}^{(3D)}_{\rm ext}(\bm{x},\bm{w})
&\equiv& \label{Uext_3d}
2N_{c}\int\,d^{3}\bar{r}
\bar{V}_{\rm ext}({\bf\bar{r}},\bar{t})
\left|\Psi({\bf\bar{r}},\bar{t})\right|^{2}\\
\bar{U}^{(3D)}_{\rm int}(\bm{x},\bm{w})
&\equiv& \label{Uint_3d}
2N_{c}
\left(\frac{1}{2}\bar{g}N\right)
\int\,d^{3}\bar{r}
\left|\Psi({\bf\bar{r}},\bar{t})\right|^{4}.
\end{eqnarray}

The derivation of the Lagrangian for the 1D case is very similar.  The major difference is the absence of a rotating--frame term.  Thus we will simply define the 1D variational potentials and present the 1D Lagrangian result.

The 1D external variational potential can be written as
\begin{eqnarray}
\bar{U}^{(1D)}_{\rm ext}(\bm{x},\bm{w})
&\equiv& 
2N_{c}\bar{L}^{(1D)}_{3}\left({\bf x},{\bf w}\right)\nonumber\\
&=&
2N_{c}\int_{-\infty}^{+\infty}\,d\bar{x}\,
\bar{V}_{\rm ext}(\bar{x},\bar{t})
\psi^{\ast}(\bar{x},\bar{t})
\psi(\bar{x},\bar{t})\nonumber\\
&=&
\sum_{j=1}^{N_{c}}
\left(
\frac{2}{\pi^{1/2}\bar{w}_{j}}
\right)
\int_{-\infty}^{+\infty}\,d\bar{x}\,\nonumber\\
&\times&
\exp
\left\{-
\frac
{\left(\bar{x}-\bar{x}_{j}\right)^{2}}
{\bar{w}_{j}^{2}}
\right\}
\bar{V}_{\rm ext}(\bar{x},\bar{t})
\label{Uext_1d}
\end{eqnarray}
Where we have used constraints 1 and 3 to simplify the integrals.  In order to apply this 1D model to a particular system, this potential must be calculated.  Derivatives of $\bar{U}_{\rm ext}$ appear in the equations of motion.

The 1D interaction variational potential is defined by
\begin{eqnarray}
\bar{U}^{(1D)}_{\rm int}(\bm{x},\bm{w})
&\equiv& 
2N_{c}\bar{L}^{(1D)}_{4}\left({\bf x},{\bf w}\right)\nonumber\\
&=&
\left(2N_{c}\right)
\frac{1}{2}\bar{g}_{1D}N
\int_{-\infty}^{\infty}\,d\bar{x}\,
\left|\psi\right|^{4}
\end{eqnarray}

We can use these definitions to write the final form of the 1D Lagrangian as follows.
\begin{eqnarray}
\bar{L}^{(1D)}
&=&
\frac{1}{N_{c}}\sum_{j=1}^{N_{c}}
\Bigg(
\dot{\bar{\alpha}}_{j}\bar{x}_{j} + 
\dot{\bar{\beta}}_{j}
\left(
\bar{x}_{j}^{2} + 
\frac{1}{2}\bar{w}_{j}^{2} 
\right) +
\frac{1}{2\bar{w}_{j}^{2}}\nonumber\\
&+& 
2\bar{\beta}_{j}^{2}\bar{w}_{j}^{2} +
\Big(
2\bar{\beta}_{j}\bar{x}_{j} +
\bar{\alpha}_{j} +
\bar{k}_{j}
\Big)^{2}
\Bigg)\nonumber\\
&+&
\frac{1}{2N_{c}}
\left(
\bar{U}^{(1D)}_{\rm ext}({\bf x},{\bf w}) + 
\bar{U}^{(1D)}_{\rm int}({\bf x},{\bf w})
\right)
\label{L_final_1d}
\end{eqnarray}
Explicit expressions for the 1D and 3D interaction potentials (which are always the same) are derived in Appendix \ref{U_int}. With the 3D and 1D Lagrangians in hand we can now derive the equations of motion for the variational parameters.

\begin{figure*}[htb]
\begin{center}
\includegraphics[width=6.75in]{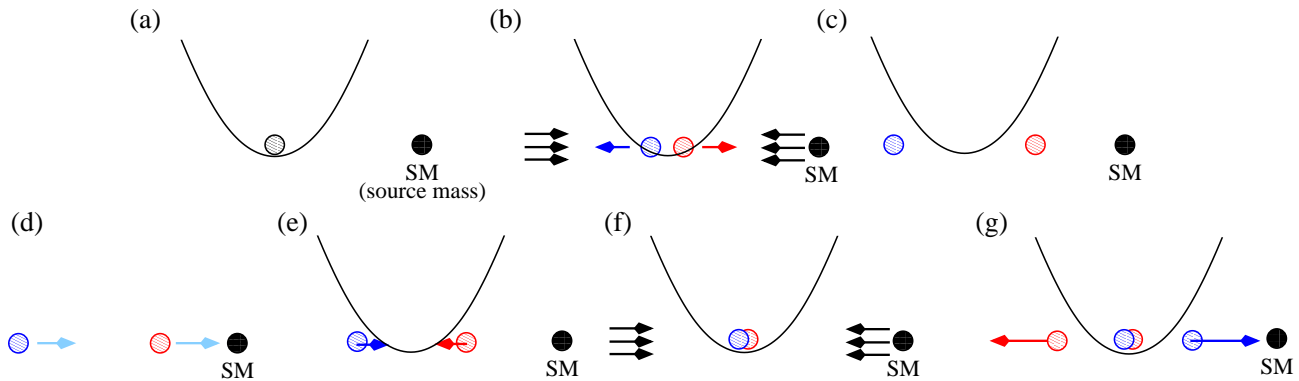}
\end{center}
\caption{(color online) 1D Atom interferometry scheme for measuring $G$. (a) A BEC is formed in the presence of an harmonic trap plus source mass (SM). (b)--(c) Initial Split Phase 1: the condensate is split by laser light and the two condensate pieces allowed to fly apart with initial speeds $\pm v$ until they stop at which point the harmonic trap is turned off. (d) Initial Split Phase 2: the two condensate pieces now experience differential gravitational accelerations due to the SM during a wait time, $T_{W}$. (e) Initial Split Phase 3: the harmonic trap is turned on pushing the condensate pieces back together.  (f)--(g) Final Split: once they overlap they are split again creating four clouds: two that fly apart with approximate speeds $\pm 2v$ and two that are motionless except for the small relative velocity developed during the wait time.}
\label{bigG_proposal}
\end{figure*} 

\subsection{LVM model equations of motion}
\label{final_eoms}

The 3D equations of motion for the Gaussian center coordinates, widths, and linear and quadratic phase parameters can obtained in two steps: (1) derive the E--L equation of motion for each variational parameter by using the 3D Lagrangian  (Eq.\,(\ref{L_final_3d})) in the standard E--L equation of motion (Eq.\,(\ref{standard_EL})) to obtain a first--order differential equation in time, (2) take the second time derivative of the equations containing $\dot{\bar{\eta}}_{j}$ and $\dot{\bar{w}}_{j\eta}$ where ($j=1,\dots,N_{c}$ and $\eta = x,y,z$) and use the other equations to eliminate the other variables from these equations.  The equations for the centers and widths together form a closed system.  

The LVM 3D equations of motion thus consist of a pair of second--order ordinary differential equation for the cloud centers and widths as well as expressions for the $\bar{\beta}_{j\eta}$ and the $\bar{\alpha}_{j\eta}$ in terms of the centers, widths and their first derivatives.  The derivation of these equations can be found in Appendix\,\ref{eoms_app}. 
\begin{subequations}
\begin{align}
\label{xj_eom_final}
\ddot{\bar{x}}_{j} &=
2\bar{\Omega}_{z}\dot{\bar{y}}_{j} +
\bar{\Omega}_{z}^{2}\bar{x}_{j} -
\frac{\partial\bar{U}^{(3D)}}{\partial\bar{x}_{j}},\\
\label{yj_eom_final}
\ddot{\bar{y}}_{j} &=
-2\bar{\Omega}_{z}\dot{\bar{x}}_{j} +
\bar{\Omega}_{z}^{2}\bar{y}_{j} -
\frac{\partial\bar{U}^{(3D)}}{\partial\bar{y}_{j}},\\
\label{zj_eom_final}
\ddot{\bar{z}}_{j} &= 
-\frac{\partial\bar{U}^{(3D)}}{\partial\bar{z}_{j}},\\
\label{wjeta_eoms_final}
\ddot{\bar{w}}_{j\eta} &=
\frac{4}{\bar{w}_{j\eta}^{3}} - 
2\frac{\partial\bar{U}^{(3D)}}
{\partial\bar{w}_{j\eta}},\\
\label{betajeta_eoms_final}
\bar{\beta}_{j\eta} &=
\frac{\dot{\bar{w}}_{j\eta}}{4\bar{w}_{j\eta}},\\
\label{alphajx_eom_final}
\bar{\alpha}_{jx} &=
\tfrac{1}{2}(\dot{\bar{x}}_{j} - \bar{\Omega}_{z}\bar{y}_{j}) - 
2\bar{\beta}_{jx}\bar{x}_{j} - \bar{k}_{jx},\\
\label{alphajy_eom_final}
\bar{\alpha}_{jy} &=
\tfrac{1}{2}(\dot{\bar{y}}_{j} + \bar{\Omega}_{z}\bar{x}_{j}) - 
2\bar{\beta}_{jy}\bar{y}_{j} - \bar{k}_{jy},\\
\label{alphajz_eom_final}
\bar{\alpha}_{jz} &=
\tfrac{1}{2}\dot{\bar{z}}_{j} - 
2\bar{\beta}_{jx}\bar{x}_{j} - \bar{k}_{jx},\\
\eta &= x,y,z\quad j=1,\dots,N_{c}\nonumber
\end{align}
\end{subequations}
In the above, $\bar{U}^{(3D)}\equiv\bar{U}^{(3D)}_{\rm ext}+\bar{U}^{(3D)}_{\rm int}$.
The equations for the cloud centers and cloud widths (Eqs.\ (\ref{xj_eom_final}), (\ref{yj_eom_final}), (\ref{zj_eom_final}), and (\ref{wjeta_eoms_final})) form a closed set that contain only the $\bar{\eta}_{j}$, $\dot{\bar{\eta}}_{j}$, $\bar{w}_{j\eta}$, and $\dot{\bar{w}}_{j\eta}$.  Once these quantities are obtained,
all of the other variational parameters can be calculated.

The equations of motion for the 1D case are derived similarly.  We will simply state the result here. They consist of a pair of second--order ordinary differential equations for the cloud centers and widths as well as expressions for 
the $\bar{\beta}_{j}$ and the $\bar{\alpha}_{j}$ in terms of the centers, widths and their time derivatives:
\begin{subequations}
\begin{align}
\label{xj_eom_final_1d}
\ddot{\bar{x}}_{j} &=
-\frac{\partial\bar{U}^{(1D)}}{\partial\bar{x}_{j}},\\
\label{wjeta_eoms_final_1d}
\ddot{\bar{w}}_{j} &=
\frac{4}{\bar{w}_{j}^{3}} - 
2\frac{\partial\bar{U}^{(1D)}}
{\partial\bar{w}_{j}},\\
\label{betajeta_eoms_final_1d}
\bar{\beta}_{j} &=
\frac{\dot{\bar{w}}_{j}}{4\bar{w}_{j}},\\
\label{alphajx_eom_final_1d}
\bar{\alpha}_{j} &=
\tfrac{1}{2}\dot{\bar{x}}_{j} - 
2\bar{\beta}_{j}\bar{x}_{j} - 
\bar{k}_{j},\\
j &=1,\dots,N_{c}\nonumber
\end{align}
\end{subequations}
Similar to the 3D case, $\bar{U}^{(1D)}\equiv\bar{U}^{(1D)}_{\rm ext}+\bar{U}^{(1D)}_{\rm int}$.
The equations for the cloud centers and cloud widths (Eqs.\ (\ref{xj_eom_final_1d}) and (\ref{wjeta_eoms_final_1d})) form a closed set that contain only the $\bar{x}_{j}$, $\dot{\bar{x}}_{j}$,$\bar{w}_{j}$, and $\dot{\bar{w}}_{j}$.  Once these quantities are obtained, all of the other variational parameters can be calculated.  

These equations of motion for the variational parameters of the 1D and 3D trial wave functions form the central result of this paper.  We note one more time that these equations hold for any external potential, $\bar{V}_{\rm ext}$.  The 1D and 3D $N_{c}$--Gaussian--cloud wave function ansatz has several areas of application including atom interferometery involving BECs and BECs moving in waveguides.

\section{AI Measurement of $G$ in Microgravity}
\label{bigG}

As an example of the use of the LVM model, we describe how it could be applied to the design of a measurement of Newton's universal gravitational constant, $G$, in a microgravity environment.  Thus we consider the idealized 1D atom--interferometry sequence shown in Fig.\ \ref{bigG_proposal} where two pieces of a BEC are separated, differentially pulled upon by a source mass (SM), recombined, and split again. In this section we describe this AI sequence in detail.

To extract the value of $G$ requires an interference pattern where only the effects of the SM are present. This can be obtained by repeating the above AI sequence many times with the SM present so that averaging over the interference patterns produced would wash out random effects of the environment. Systematic effects of the environment as well as the effect of the SM would still be present in the averaged pattern.  To remove systematic effects of the environment the SM would then be taken away and the AI sequence repeated many more times and these SM--absent interference patterns would again be averaged.  The difference between the averaged SM--present and SM-absent patterns would leave a pattern in which only the effects of the SM are present.  The pattern due to the SM where environmental effects are neglected can be approximately simulated by the variational method described here.

\begin{figure*}[tb]
\begin{center}
\includegraphics[width=7.00in]{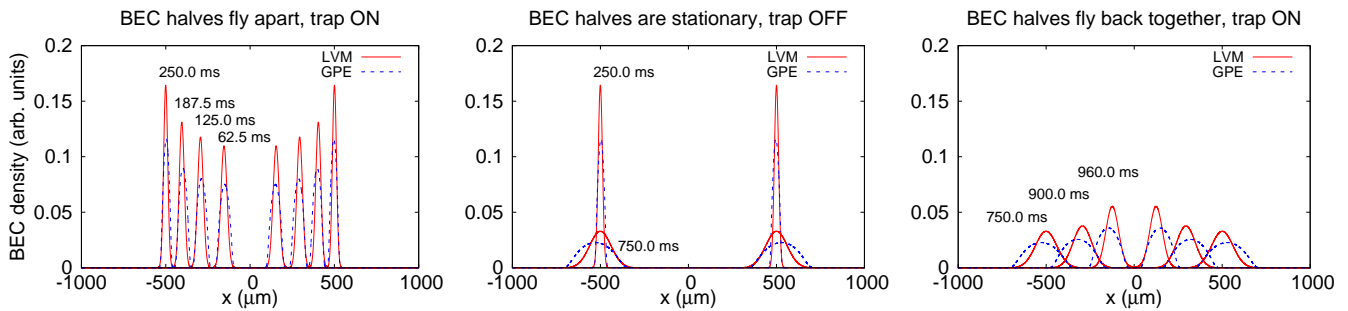}
\end{center}
\caption{(color online) Comparison of the evolution of the condensate density during the ``initial split'' phase of the AI scheme (see Fig.~\ref{bigG_proposal} panels (a)--(f)).  The left panel shows the LVM (red, solid lines) and GPE (blue, dashed lines) results at four different times as the condensate pieces are flying apart and the trap is on. Note that each curve consists of two symmetically placed peaks and only the left peak is given a time label. The middle panel shows their evolution at the beginning and end of the period when the trap is off. The right panel shows the evolution after the trap is turned back on and the condensate pieces re--overlap.  The conditions for this case are $N=10^{4}$\,$^{87}$Rb atoms, the trap potential frequency is $\omega_{T}/2\pi =1$\,Hz, the wait time is $T_{W}=500$\,ms., the initial velocity is $v=3.14\times10^{-3}$ m/s and there is no source mass present.}
\label{lvm_gpe_cmp}
\end{figure*}

The basic idea of the AI sequence is that the relative velocity developed between the two condensate pieces due to the differential acceleration produced by the gravity of the source mass. This relative velocity is imprinted on the final--state interference pattern since the velocity distribution is proportional to the condensate phase gradient. 

The value of $G$ can then be extracted from the interference data.  This sequence is imagined as the essential step in an AI measurement of $G$. It is assumed to be carried out in a microgravity environment such as NASA's Cold Atom Laboratory aboard the International Space Station.  A more sophisticated version of this sequence would be required for a precision measurement of $G$ under such conditions.  

The basic sequence begins with a condensate formed at the center of a 1D harmonic trap with frequency $\omega_{T}$ with a SM present (Fig.\,\ref{bigG_proposal}(a)).  The overall sequence consists of an ``initial split'' where this initial condensate is separated into two pieces which are then rejoined and a ``final split'' where the rejoined pieces are split again and the interference pattern is imaged.
The initial split (IS) has three phases as we now describe.

The speed $v$ imparted to the condensate pieces is assumed in this work to come from applying identical counterpropagating laser beams to the condensate.  An atom will receive a momentum kick by absorbing a photon from one beam and emitting it into the other beam.  The momentum imparted to a condensate atom is $2\hbar k_{L}$ where $k_{L}=2\pi/\lambda_{L}$ where $\lambda_{L}$ is the wavelength of the laser.  Laser pulses can be engineered so that this basic momentum--kick process happens $m$ times so that the velocity kick the atoms receive is $v=4\pi\hbar m/(M\lambda_{L})$ where $M$ is the mass of the condensate atom.  In the examples given below we assume that $m=8$ so that, for $^{87}$Rb atoms, $v=9.14\times10^{2}$\,m/s.

In IS phase 1 the condensate is separated into two equal pieces by a sequence of optical--lattice pulses~\cite{PhysRevA.82.063613} the total effect of which is to change the velocity of one condensate piece by $+v$ and the other by $-v$ (Fig.\,\ref{bigG_proposal}(b)).  The initially motionless condensate pieces thus fly apart with velocities $-v$ (blue circle to the left) and $+v$ (red circle to the right) and both come to a stop at the turning points of the harmonic potential after one quarter of the trap period.  We call this time $t=T_{1/4}$ and is shown in Fig.\,\ref{bigG_proposal}(c).

In IS phase 2 the harmonic trap is turned off at time $t=T_{1/4}$ ideally leaving the two pieces motionless. The system is then allowed to evolve during the time interval $T_{1/4}\le t \le T_{1/4}+T_{\rm w}$.  We call this evolution time, $T_{\rm w}$, the ``wait period''.  This is shown in Fig.\,\ref{bigG_proposal}(d). During this time the gravitational pull of the SM causes a differential acceleration of the two condensate pieces since one piece is further away from the SM than the other.  This differential acceleration causes the two pieces to develop a relative velocity. 

In IS phase 3 the trap is then turned back on which brings the condensate pieces back together during the time interval $T_{1/4}+T_{\rm w}\le t\le 2T_{1/4}+T_{\rm w}$ as shown in  Fig.\,\ref{bigG_proposal}(e).  Once they overlap again the blue (left) piece will be moving at velocity $+v$ plus a small difference due to the SM and the red (right) piece will be moving at $-v$ plus a small difference.  The initial split takes place during the time interval $0\le t\le 2T_{1/4}+T_{\rm w}\equiv T_{\rm is}$ and shown in Figs.\,\ref{bigG_proposal}(a)--(e).

Once the two pieces are overlapped again the same set of optical--lattice pulses is applied as before (Fig.\,\ref{bigG_proposal}(f)).  This causes the red condensate piece to split in two with a piece moving at velocity approximately $-2v$ and the other at zero velocity plus a small amount due to the pull of the SM.  The blue half is also split into a piece moving at approximate velocity $+2v$ and the other approximately motionless except again for a small deviation. These two nearly motionless pieces will remain overlapped and will have the small relative velocity that was developed during the wait time when the trap was turned off. We assume that the wait time is much longer than a quarter period of the harmonic trap ($T_{\rm w}\gg T_{1/4}$).  

Finally, the two fast condensate pieces will fly away leaving the nearly motionless pieces behind as shown in Fig.\,\ref{bigG_proposal}(g).  We will refer to this sequence, shown in Figs.\,\ref{bigG_proposal}(f)--(g), as the ``final split''.  Imaging this middle cloud will leave an interference pattern due to the relative velocity of the two condensate pieces and this pattern will be due only to the differential gravitational pull of the SM during the wait time $T_{W}$.  

\section{Applying the LVM model to the AI sequence}
\label{lvm_analysis}

\subsection{Comparison of the LVM model and GPE solutions}
\label{lvm_gpe_compare}

In this section we show how the LVM model can be used to analyze the atom interferometry sequence described above.  We emphasize here that the LVM model is not sufficiently accurate to be used in the analysis of a {\em precision measurement} rather it is to be used in the preliminary design of the experiment.  To show this, we simulated the evolution of the condensate wave function during the initial split of this AI sequence using the 1D time--dependent GPE and the 1D LVM model described earlier.  The comparison is shown in Fig.\,\ref{lvm_gpe_cmp}. The details of how the simulation was conducted can be found in Appendix~\ref{gpe_sim}.  

This comparison shows that, while the Gaussian--approximate trial wave functions only qualitatively agree their GPE counterparts, the motion of the GPE and LVM wave packet centers and widths track very well.  The variational wave function will only agree with the exact solution to the extent that variational wave function can be modified to fit the exact solution by varying its parameters. 

Modeling performed for the final design of a precision AI measurement will require the numerical solution of the full 3D Gross--Pitaevskii equation.  Since experimental conditions for precision measurements are typically extreme, numerical solution of the 3D GPE will require a substantial effort.  This effort usually makes solving the 3D GPE too expensive for the purposes of preliminary AI design.  We shall elaborate on this point below. 

The purpose of the LVM model is to facilitate the {\em preliminary} design of a precision measurement. It can serve this purpose in several ways. First, it can provide a simple and physically intuitive picture of how the measured quantity can be extracted from the experimental data.  In some cases an approximate analytical expression for the measured quantity in terms of experimental parameters can be derived. Second, the model can be used to estimate the sensitivity of the measurement.  Finally, this expression can be used to speed up significantly the time spent on the preliminary design of an AI sequence by providing estimates of the values of these parameters thus constraining the size of the experimental parameter space. We illustrate these features of the LVM model by applying it to the AI sequence described above.

\subsection{Approximating the interference pattern using the LVM model}
\label{lvm_ip}

The equations of motion for the case when an harmonic trap with frequency $\omega_{T}$ is present along with a source mass of mass $M_{SM}$ and located at $x_{SM}$ can be found by computing the derivatives found in Eqs.\ (\ref{xj_eom_final_1d}) and (\ref{wjeta_eoms_final_1d}).  The external varational potential for this case, given by Eq.\ (\ref{u_ext_final_1d}), is derived in Appendix \ref{U_ext} and the interaction variational potential is given in Eq.\ (\ref{u_int_final_1d}) of Appendix \ref{U_int}. 

The resulting equations of motion have a striking similarity to equations of motion produced by  Newton's second law.  They can be written as follows:
\begin{subequations}
\begin{align}
\label{xj_eom_specific}
\ddot{\bar{x}}_{j} +
\left(
\bar{\omega}_{T}^{2} -
2\bar{\omega}_{SM}^{2}
\right)\bar{x}_{j}
&= 
\bar{\omega}_{SM}^{2}\bar{x}_{SM} -
X_{j}({\bf x},{\bf w}),\\
\label{wj_eom_specific}
\ddot{\bar{w}}_{j} +
\left(
\bar{\omega}_{T}^{2} -
2\bar{\omega}_{SM}^{2}
\right)\bar{w}_{j}
&=
\frac{4}{\bar{w}_{j}^{3}} +
\frac{\bar{g}N}{(2\pi)^{1/2}\bar{w}_{j}^{2}} -
W_{j}({\bf x},{\bf w})\nonumber\\
j &=1,2.
\end{align}
\end{subequations}
The terms $X_{j}({\bf x},{\bf w})$ and $W_{j}({\bf x},{\bf w})$ represent ``repulsion forces'' due to the overlap of different clouds and are given by
\begin{eqnarray}
X_{j}({\bf x},{\bf w}) &=&
\left(
\frac{4\bar{g}N}{\pi^{1/2}}
\right)\\
&\times&
\sum_{j_{1}\ne j}^{2}
\left(
\frac
{\left(\bar{x}_{j_{1}}-\bar{x}_{j}\right)
\exp
\left\{
-\frac
{\left(\bar{x}_{j_{1}}-\bar{x}_{j}\right)^{2}}
{\bar{w}_{j_{1}}^{2}+\bar{w}_{j}^{2}}
\right\}
}
{
\left(\bar{w}_{j_{1}}^{2}+\bar{w}_{j}^{2}\right)^{3/2}
}
\right)\nonumber
\label{x_overlap}
\end{eqnarray}
and
\begin{eqnarray}
W_{j}({\bf x},{\bf w}) &=&
\left(
\frac{4\bar{g}N}{\pi^{1/2}}
\right)
\sum_{j_{1}\ne j}^{2}
\left(
\frac
{\bar{w}_{j}
\exp
\left\{
-\frac
{\left(\bar{x}_{j_{1}}-\bar{x}_{j}\right)^{2}}
{\bar{w}_{j_{1}}^{2}+\bar{w}_{j}^{2}}
\right\}
}
{
\left(\bar{w}_{j_{1}}^{2}+\bar{w}_{j}^{2}\right)^{5/2}
}
\right)\nonumber\\
&\times&
\left[
2\left(\bar{x}_{j_{1}}-\bar{x}_{j}\right)^{2} -
\left(\bar{w}_{j_{1}}^{2}+\bar{w}_{j}^{2}\right)
\right],
\end{eqnarray}
where $j=1,2$.  

The LVM model can provide an approximate expression for the interference pattern produced by the above AI sequence.  This can be done in two steps: (1) use the 1D trial wave function with $N_{c}=2$ from Eq.\,(\ref{1d_trial_wf}) to derive the condensate density at the end of the sequence, and (2) use approximate analytical solutions to the equations of motion given by Eqs.\,(\ref{xj_eom_specific}) and (\ref{wj_eom_specific}) to obtain expressions for the variational parameters $\bar{x}_{1}$, $\dot{\bar{x}}_{1}$, $\bar{x}_{2}$, $\dot{\bar{x}}_{2}$, $\bar{w}_{1}$, $\dot{\bar{w}}_{1}$, $\bar{w}_{2}$, and $\dot{\bar{w}}_{2}$ at the end of the sequence. 

The condensate trial wave function at the end of the initial split ($t=T_{\rm is}$) for two clouds ($N_{c}=2$) is given by Eq.\,(\ref{1d_trial_wf}) and can be written as follows:
\begin{eqnarray}
\psi_{\rm is}(\bar{x},\bar{T}_{\rm is}) 
&=& 
(2\pi^{1/2}\bar{w}_{1})^{-1/2}
e^{-\frac
{\left(\bar{x}-\bar{x}_{1}\right)^{2}}
{2\bar{w}_{1}^{2}} +
i((\tfrac{1}{2}\dot{\bar{x}}_{1}-2\bar{\beta}_{1}\bar{x}_{1})\bar{x} + \bar{\beta}_{1}\bar{x}^{2})
}\nonumber\\
&+& 
(2\pi^{1/2}\bar{w}_{2})^{-1/2}
e^{-\frac
{\left(\bar{x}-\bar{x}_{2})\right)^{2}}
{2\bar{w}_{2}^{2}} +
i((\tfrac{1}{2}\dot{\bar{x}}_{1}-2\bar{\beta}_{2}\bar{x}_{2})\bar{x} + \bar{\beta}_{2}\bar{x}^{2})
}\nonumber\\
&\equiv&
\frac{1}{\sqrt{2}}
\left(
\psi_{{\rm is},1}(\bar{x},\bar{T}_{\rm is}) + 
\psi_{{\rm is},2}(\bar{x},\bar{T}_{\rm is})
\right)
\end{eqnarray}
where we have used Eq.\,(\ref{alphajx_eom_final_1d}). All variational parameters above are evaluated at $t=T_{\rm is}$.  Since $\bar{\beta}_{j}= \dot{\bar{w}}_{j}/(4\bar{w}_{j})$ (from Eq.\,(\ref{betajeta_eoms_final_1d}) the wave function is expressed entirely in terms of the $\bar{x}_{j}$, $\bar{w}_{j}$ and their derivatives.  These quantities can be found by solving Eqs.\,(\ref{xj_eom_specific}) and (\ref{wj_eom_specific}).

We can model the effect of the second splitting by multiplying $\psi(\bar{x},\bar{T}_{\rm is})$ by a factor which splits each exisiting cloud into two clouds: one which is boosted by $v$ and one boosted by $-v$.  Thus, in scaled units, the final--split wave function at $t=T_{\rm is}$ becomes
\begin{eqnarray}
\psi_{\rm fs}(\bar{x},\bar{T}_{\rm is}) 
&=& 
\frac{1}{2}
\left(
e^{+i\bar{v}\bar{x}/2} + e^{-i\bar{v}\bar{x}/2}
\right)
\left(
\psi_{{\rm is},1} + \psi_{{\rm is},2}
\right)\nonumber\\
&=&
\frac{1}{2}e^{-i\bar{v}\bar{x}/2}\psi_{{\rm is},1} +
\frac{1}{2}e^{+i\bar{v}\bar{x}/2}\psi_{{\rm is},1}\nonumber\\
&+&
\frac{1}{2}e^{-i\bar{v}\bar{x}/2}\psi_{{\rm is},2} +
\frac{1}{2}e^{+i\bar{v}\bar{x}/2}\psi_{{\rm is},2}\nonumber\\
&\equiv&
\psi_{1,1} + \psi_{1,2} + \psi_{2,1} + \psi_{2,2}.
\end{eqnarray}
The final--split wave function consists of four clouds.  The velocities of these clouds just after the second splitting are $\approx 0$ for clouds labeled (1,1) and (2,2), $\approx +2v$ for cloud (1,2) and $\approx -2v$ for cloud (2,1). 

Clouds (1,2) and (2,1) fly rapidly away from the center and after a short time, $\tau$, the overlap of the fast clouds with those at the center can be neglected.  At this time in the AI sequence the density of the condensate can be imaged to obtain an interference pattern. To obtain a simple approximation for the condensate density at this time we assume that the change in the variational parameters during the time $\tau$ can be neglected and we approximate their values at time $t=T_{\rm is}+\tau$ with their values at $t=T_{\rm is}$.  

The condensate density at the trap center can be found from the squared modulus of the sum of the wave functions of the two nearly motionless clouds. The result for the condensate density near the trap center after the final split is (in SI units)
\begin{eqnarray}
\rho(x,T_{\rm is}) 
&=& 
\left|\psi_{1,1} + \psi_{2,2}\right|^{2}\nonumber\\
&=&
\tfrac{1}{4}(\pi^{\tfrac{1}{2}}w_{1})^{-1}
e^{-(x-x_{1})^{2}/w_{1}^{2}}\\
&+&
\tfrac{1}{4}(\pi^{\tfrac{1}{2}}w_{2})^{-1}
e^{-(x-x_{2})^{2}/w_{2}^{2}}\nonumber\\
&+&
\tfrac{1}{2}
(\pi^{\tfrac{1}{2}}w_{1})^{-\tfrac{1}{2}}
(\pi^{\tfrac{1}{2}}w_{2})^{-\tfrac{1}{2}}\nonumber\\
&\times&
e^{
-(x-x_{1})^{2}/(2w_{1}^{2})
-(x-x_{2})^{2}/(2w_{2}^{2})
}
\cos
\left(
\phi(x)
\right),
\label{int_pattern}\nonumber
\end{eqnarray}
where
\begin{equation}
\phi(x)=k_{f}x+(\beta_{1}-\beta_{2})x^{2}
\label{bec_phase}
\end{equation}
is the final condensate phase difference and 
\begin{equation}
k_{f} =
\frac{M}{\hbar}
\left(
\dot{x}_{2}-\dot{x}_{1}
\right) + v +
2\beta_{1}x_{1} - 2\beta_{2}x_{2}.
\label{int_freq}
\end{equation}
The quantity $k_{f}$ is the spatial frequency near the center of the interference pattern.  We note that these variational quantities are understood to be evaluated at $t=T_{\rm is}$.  We can now find an approximate interference pattern by solving Eqs.\,(\ref{xj_eom_specific}) and (\ref{wj_eom_specific}) for the variational parameters either numerically or by finding approximate analytical solutions.  

\begin{figure}[tb]
\begin{center}
\includegraphics[height=3.50in,angle=-90]{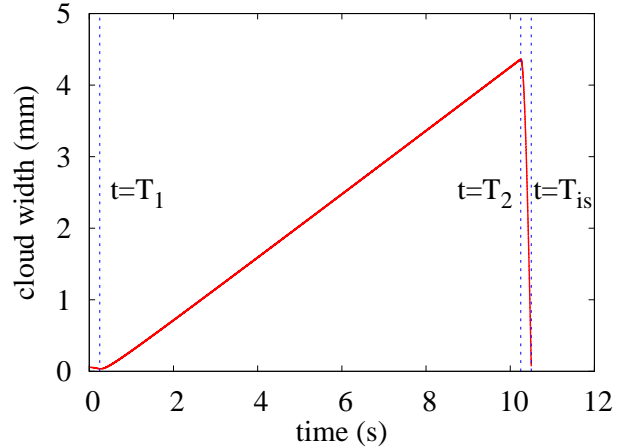}
\end{center}
\caption{(color online) Cloud width versus time, $w_{1}(t)$, during the IS from the numerical solution of the full system of equations, (\ref{xj_eom_specific}) and (\ref{wj_eom_specific}).  The conditions are $M_{SM}=100$ kg, $x_{SM}=0.1$ m, $v=9.14\times 10^{-2}$ m/s, $\omega_{T}/2\pi=1$ Hz, and $T_{W}=10$ seconds.  The vertical dotted lines indicate the end--times of the three phases of the IS.}
\label{width_fig}
\end{figure} 

The presence of a source mass has two major effects on the interference pattern: (1) it creates interference fringes and (2) the peak of the pattern is shifted away from the zero--mass position towards the source--mass. We will call the width of the central fringe, $\lambda_{f}$, and the shift of the peak due to the source mass is called, $x_{c}$.  Later we will show how $G$ can be obtained from the experimental parameters and either of these two characteristics of the interference pattern. 

The final condensate phase, $\phi(x)$, predicts an interference pattern that displays equally spaced fringes near the center of the pattern due to the linear term. The fringe spacing narrows away from the center due to the quadratic term.  The gradient of the phase  measures the local relative velocity of the overlapping clouds. 

This relative velocity is caused by (1) the differential accelerations of the two clouds during the wait period and (2) the changing widths of the two clouds due to condensate atom--atom interactions.  Interactions also cause the fringe spacing to narrow away from the center of the interference pattern.  The $\beta$ parameters in the LVM model describe how the clouds expand and/or contract due to the competition between the repulsive interactions and the harmonic confinement.  They also account for any shift, due to atom--atom interactions, in the value of $G$ inferred from the interference pattern.

\subsection{Extracting $G$ from the inteference pattern}
\label{bigG_value}

In modeling a {\em precision measurement} the value of $G$ would be inferred from repeated GPE simulations of the experiment using different values of $G$ until a satisfactory match between the measured and simulated interference patterns was obtained.  This is a necessary (but very expensive, see below) procedure.  Such repeated simulations with different $G$ values can also be performed using the LVM model even though its accuracy is not sufficient to make a final decision about any particular AI scheme. 

The LVM model can also help in designing an experiment by providing an approximate expression for $G$ in terms of the experimental parameters and characteristics of the interference pattern. This can be done by approximating the solution of the LVM equations of motion for each step of the proposed AI sequence. We now show this by deriving expressions for $G$ for the AI sequence described in Section \ref{bigG}.

\begin{figure*}[htb]
\begin{center}
\includegraphics[width=7.00in]{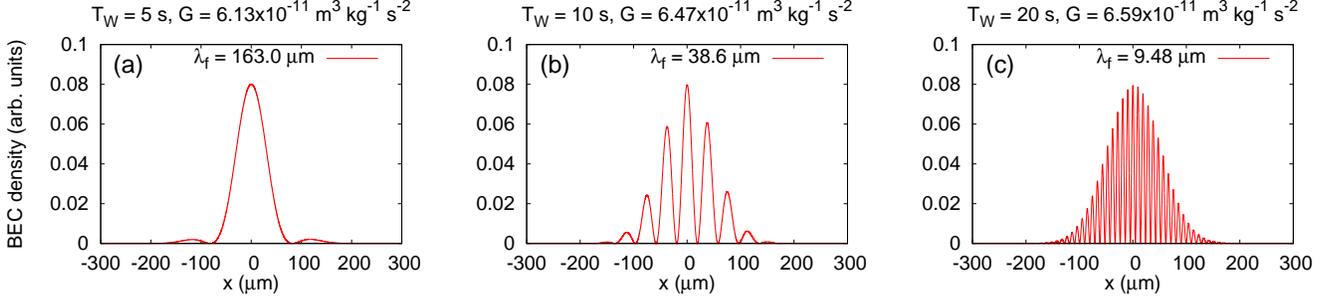}
\end{center}
\caption{(color online) Interference patterns for three different wait times showing how the width of the central fringe decreases as $T_{W}$ increases. (a) $T_{W}=5$ s, $\lambda_{f}=163.0\,\mu$m (b) $T_{W}=10$ s, $\lambda_{f}=38.6\,\mu$m and (c) $T_{W}=20$ s, $\lambda_{f}=9.48\,\mu$m. Except for the wait times the conditions are the same as in Fig.\,\ref{width_fig}: $M_{SM}=100$ kg, $x_{SM}=0.1$ m, $v=9.14\times 10^{-2}$ m/s, $\omega_{T}/2\pi=1$ Hz.  The resulting values for $G$ are (a) $G = 6.13\times10^{-11}$ m$^{3}$\,kg$^{-1}$\,s$^{-2}$, (b) $G = 6.47\times10^{-11}$ m$^{3}$\,kg$^{-1}$\,s$^{-2}$, and (c) $G = 6.59\times10^{-11}$ m$^{3}$\,kg$^{-1}$\,s$^{-2}$.}
\label{ip_fig}
\end{figure*} 

To find approximate solutions of Eqs.\,(\ref{xj_eom_specific}) and (\ref{wj_eom_specific}) we will make three assumptions: (1) the gravity effect on the condensate is small compared to the harmonic confinement so that $\omega_{SM}\ll\omega_{T}$ (see Eq.\,(\ref{gf})), (2) the overlap terms, $X_{j}$ and $W_{j}$, can be neglected, and (3) the condensate width at the end of IS phase 2 can be approximated by the product of the IS phase 2 width expansion rate and the wait time, i.e., $\bar{w}(\bar{T}_{2})\approx\dot{\bar{w}}(\bar{T}_{2})\bar{T}_{W}$.  

One immediate consequence of this is that we can neglect $\omega_{SM}$ in the equations of motion for the $\bar{w}_{j}$ and so they are approximately unaffected by gravity during the entire AI sequence.  With these approximations we can also infer that the time evolution of the widths of the two clouds are approximately the same, i.e., $\bar{w}_{1}(\bar{t}) \approx \bar{w}_{2}(\bar{t})$ and also $\dot{\bar{w}}_{1}(\bar{t}) \approx \dot{\bar{w}}_{2}(\bar{t})$.  This also implies that $\bar{\beta}_{1}(\bar{t})\approx\bar{\beta}_{2}(\bar{t})$.

These approximations also enable us to find analytical solutions of  Eq.\,(\ref{xj_eom_specific}) for the cloud center positions in each of the three phases of the initial split (panels (c), (d), and (e) of Fig.\,\ref{bigG_proposal}, respectively).  In phases 1 and 3 the solutions are $\sin(\bar{\omega}_{T}\bar{t})$ and $\cos(\bar{\omega}_{T}\bar{t})$ and, in phase 2, the solutions are $\sinh(\bar{\omega}_{SM}\bar{t})$ and $\cosh(\bar{\omega}_{SM}\bar{t})$.  Using the initial conditions $\bar{x}_{1}(0)=\bar{x}_{2}(0)=0$, $\dot{\bar{x}}_{1}(0)=-\bar{v}$, and $\dot{\bar{x}}_{2}(0)=+\bar{v}$ we can solve equations of motion for the center coordinates in each IS phase using the final values of $\bar{x}_{j}$ and $\dot{\bar{x}}_{j}$ in the previous phase as the initial values in the next phase.

This straightforward procedure enables us to obtain approximate expressions for the center positions and velocities at $t=T_{is}$.  The results are
\begin{eqnarray}
\bar{x}_{j}(\bar{T}_{\rm is})
&=&
\frac{\sqrt{2}\bar{\omega}_{SM}}{\bar{\omega}_{T}}
\left(
\tfrac{1}{2}\bar{x}_{SM}+\epsilon_{j}\frac{\bar{v}}{\bar{\omega}_{T}}
\right)
\sinh(\phi_{SM})\\
\dot{\bar{x}}_{j}(\bar{T}_{\rm is})
&=&
\bar{\omega}_{T}
\left[
\tfrac{1}{2}\bar{x}_{SM} -
\left(
\tfrac{1}{2}\bar{x}_{SM}+\epsilon_{j}\frac{\bar{v}}{\bar{\omega}_{T}}
\right)
\cosh(\phi_{SM})
\right],\nonumber
\label{acv} 
\end{eqnarray}
where $j=1,2$, $\epsilon_{1}=-1$, $\epsilon_{2}=+1$, $\phi_{SM}\equiv\sqrt{2}\omega_{SM}T_{W}$ and from Eq.\,(\ref{grav_freq})
\begin{equation}
\omega_{SM}=\left(\frac{GM_{SM}}{\left|x_{SM}\right|^{3}}\right)^{1/2}.
\label{gf}
\end{equation}

It is also possible to find an approximate value for $\bar{\beta}_{1}(\bar{T}_{is})$ (assumed to be equal to $\bar{\beta}_{2}(\bar{T}_{is})$) by approximating the width equation of motion (Eq.\,(\ref{wj_eom_specific})) during phase 3 of the IS.  An example of how the width behaves during this time is shown in Fig.\,\ref{width_fig} between the two rightmost vertical dotted lines.  The cloud width is initially quite large and drops rapidly to a value close to its starting value.  

Thus during the IS phase 3 period all of the interaction terms in Eq.\,(\ref{wj_eom_specific}) can be neglected and thus the equation of motion can be approximated as (for cloud 1)
\begin{equation}
\ddot{\bar{w}}_{1}+\bar{\omega}_{T}^{2}\bar{w}_{1} \approx 0,
\end{equation}
with initial conditions $\bar{w}_{1}(\bar{T}_{2})$ and $\dot{\bar{w}}_{1}(\bar{T}_{2})$ at $\bar{t}=\bar{T}_{2}$.  This can be easily solved and, since the duration of phase 3 is one--quarter of the trap period, we can write $\bar{w}_{1}(\bar{T}_{is})$ and $\dot{\bar{w}}_{1}(\bar{T}_{is})$ in terms of $\bar{w}_{1}(\bar{T}_{2})$ and $\dot{\bar{w}}_{1}(\bar{T}_{2})$ as follows
\begin{equation}
\bar{w}_{1}(\bar{T}_{is}) = 
\frac{\dot{\bar{w}}_{1}(\bar{T}_{2})}{\bar{\omega}_{T}}
\end{equation}
and
\begin{equation}
\dot{\bar{w}}_{1}(\bar{T}_{is}) = 
-\bar{w}_{T}\bar{w}_{1}(\bar{T}_{2}) \approx
-\bar{w}_{T}\dot{\bar{w}}_{1}(\bar{T}_{2})\bar{T}_{W},
\end{equation}
where in the last equality we have used the approximation mentioned above that the cloud width at the end of IS phase 2 is approximately the width velocity at $t=T_{2}$ multiplied by the wait time, $T_{W}$.  So, finally we have an approximate expression for the $\beta$ parameters at $t=T_{is}$:
\begin{equation}
\bar{\beta}_{2}(\bar{T}_{is}) \approx
\bar{\beta}_{1}(\bar{T}_{is}) =
\frac{\dot{\bar{w}}_{1}(\bar{T}_{is})}{4\bar{w}_{1}(\bar{T}_{is})} \approx
-\frac{1}{4}\bar{\omega}_{T}^{2}\bar{T}_{W}.
\label{approx_beta}
\end{equation}
This expression enables us to derive an expression for $G$ solely in terms of the experimental parameters and characteristics of the measured interference pattern.

We can use the expressions for the $\bar{x}_{j}(\bar{T}_{is})$ and the $\dot{\bar{x}}_{j}(\bar{T}_{is})$ from Eqs.\,(\ref{acv})) and the $\bar{\beta}_{j}(\bar{T}_{is})$ in Eq.\,(\ref{approx_beta}) to approximate $k_{f}$ in Eq.\,(\ref{int_freq}). This equation can be written as (in SI units)
\begin{equation}
\frac{\hbar k_{f}}{2Mv} = \frac{\pi\hbar}{Mv\lambda_{f}} = 
1 + \phi_{SM}\sinh(\phi_{SM}) - \cosh(\phi_{SM}).
\label{geq1}
\end{equation}
The left--hand--side (LHS) of this equation contains only experimental parameters ($M$ and $v$) and a characteristic of the interference pattern ($\lambda_{f}$ is the width of the central fringe).  The value of the LHS can thus be used to find the value of $\phi_{SM}=\sqrt{2}\omega_{SM}T_{W}$ from which $G$ can be obtained.  Thus this expression can be used to estimate the value of $G$ from the experimental parameters and the interference pattern.

This is most easily seen if we assume that $\phi_{SM}\ll 1$ in which case the above expression for $G$ can be further approximated as $k_{f}\approx Mv\phi_{SM}^{2}/2\hbar$. So, in SI units, we can write
\begin{eqnarray}
G 
&\approx& \left(\frac{\hbar x_{SM}^{3}}{2MvM_{SM}T_{W}^{2}}\right)k_{f}\nonumber\\
&=& \left(\frac{\hbar x_{SM}^{3}}{2MvM_{SM}T_{W}^{2}}\right)
\left(\frac{2\pi}{\lambda_{f}}\right)
\label{G_from_lf}
\end{eqnarray}
where $M$ is the mass of a condensate atom and $\lambda_{f}$ is the width of the central fringe of the interference pattern. 

This formula expresses $G$ in terms of the experimental parameters, $T_{W}$, $x_{SM}$, $M_{SM}$, and $v$, and, $\lambda_{f}$, the width of the central fringe of the interference pattern.  This quantity is easily inferred from the pattern. Interference patterns for several sets of experimental parameters are shown in Fig.\,\ref{ip_fig}.  In particular, the figure shows how the patterns change as the wait time, $T_{W}$, is increased.  The figure also shows the values of $G$ given by the above formula for each pattern.

It is also possible to write an expression for $G$ in terms of the shift, $x_{c}$, of the interference pattern maximum away from the maximum of the pattern obtained when the source mass is absent.  The result is
\begin{eqnarray}
x_{c} 
&\approx&
\tfrac{1}{2}
\left(x_{1}+x_{2}\right)\nonumber\\
x_{c} 
&=&
\frac{x_{SM}}{2\omega_{T}T_{W}}\phi_{SM}\sinh(\phi_{SM})\nonumber\\
x_{c} 
&\approx&
\frac{x_{SM}}{2\omega_{T}T_{W}}\phi_{SM}^{2},
\quad
\phi_{SM}\ll 1\ {\rm so\ that}\nonumber\\
G
&\approx&
\left(\frac{x_{SM}^{2}\omega_{T}}{T_{W}M_{SM}}\right)x_{c}.
\label{G_from_xc}
\end{eqnarray}
Equations (\ref{G_from_lf}) and (\ref{G_from_xc}) display the ability of LVM model to show how $G$ can be obtained from the experimental parameters and the data.  They can further be used to estimate the sensitivity of the measured $G$ value to each of the experimental parameters and to develop a preliminary error budget for the experiment.
\subsection{Using the LVM model for preliminary AI design}
\label{prelim}

One of the features of the LVM model is that different AI schemes can be rapidly evaluated and compared. The scheme presented above only splits the condensate into two fast pieces. Other possible AI schemes might, for example, split the condensate into three, five or more pieces (see Ref.\,\cite{PhysRevA.82.063613}).  Whether such ideas are useful can only be determined by modeling the particular AI sequence envisioned.  In general this will require many iterations in order to come to a conclusion about the candidate AI sequence.  We have also shown above that the model is also useful in shrinking the experimental parameter space.

If this modeling is conducted by numerical solution of the GPE, this will require many runs each of which is very expensive.  For example, in the simulation shown in Fig.\,\ref{lvm_gpe_cmp} the GPE run needed $10^{4}$ seconds to finish using a fast desktop computer while the LVM simulation required $< 1$ second to finish.  This simulation where $T_{W} = 0.5$\,s and $v = 3.15\times10^{-3}$\,m/s was very different from the much more extreme conditions found in Fig.\,\ref{ip_fig}(c) where $T_{W}=20$\,s and $v=9.14\times10^{-2}$\,m/s. The size of the grid box in the second simulation was 30 times larger and the evolution time was 40 times longer making the full simulation take 1200 times longer.  Thus, if the same GPE simulator used for the first simulation were used to perform the second simulation, it would have taken the desktop computer more than $10^{7}$ seconds (or about 4 months) to finish this single simulation.  The LVM simulation took less than three minutes. If dozens or hundreds of preliminary modeling runs were needed, it would clearly not be practical to use the GPE even if such modeling were done on a supercomputer.  Furthermore, a realistic AI experiment for measuring $G$ would take place in three dimensions and would require 3D modeling.  This would make simulation time for the GPE even longer but would not be significantly longer for the LVM model.

We believe that it is not necessary to use the GPE for {\em preliminary design} modeling of candidate AI sequences.  This can be accomplished with the LVM and {\em precision modeling} can carried out using the GPE only for a small number of likely AI candidate schemes.  A number of such likely 3D schemes will be the subject of a future publication.

\section{Summary}
\label{summary}

In this paper we presented a variational technique for approximating the solution of the time--dependent Gross--Pitaevskii equation,  for situations where the condensate is split into multiple parts as is common in atom interferometry processes.  The technique presented can approximate the solution of either the one--dimensional GPE or the three--dimensional GPE in the laboratory or rotating frame. 

Each part of the condensate is modeled as a Gaussian with time--dependent center, width, and linear and quadratic phase parameters that enable the centers and widths to change. Different parts of the condensate are allowed to overlap but exchange of atoms between different condensate parts is assumed to be negligible.  The resulting equations of motion for the center coordinates, widths, and phase parameters is a system of ordinary differential equations in time that can be rapidly solved.  

The power of this model is that this set of equations is cast in terms of derivatives of an arbitrary external potential and thus can be applied to many different AI schemes.  This model should enable rapid prototyping and design of novel AI schemes for applications especially in microgravity environments.

\begin{acknowledgments}
This material is based upon work supported by the U.S.\ National Science Foundation  under grant numbers  PHY--1004975, PHY--0758111, the Physics Frontier Center grant PHY--0822671 and by the National Institute of Standards and Technology.
\end{acknowledgments}

\clearpage
\bibliography{tools_designing_ai_microgravity}{}
\clearpage
\appendix
\section{3D LVM Lagrangian}
\label{3d_LVM_Lagrangian}

Here we sketch the derivation of the 3D Lagrangian.  This Lagrangian has five terms, defined in Eq.\ (\ref{3d_lag_int}), and we will derive $\bar{L}^{(3D)}_{1}$, $\bar{L}^{(3D)}_{2}$ and $\bar{L}^{(3D)}_{5}$ explicitly. Further we will show that $\bar{L}^{(3D)}_{3}$ and $\bar{L}^{(3D)}_{4}$ only depend on the $\bm{x}$ and $\bm{w}$ variational parameters. This in turn motivates the introduction of the ``variational'' potentials, $\bar{U}_{\rm ext}(\bm{x},\bm{w})$ and $\bar{U}_{\rm int}(\bm{x},\bm{w})$, which later appear in the equations of motion for the Gaussian center coordinates and widths.  The explicit expression for $\bar{U}_{\rm int}(\bm{x},\bm{w})$ is given in Appendix \ref{U_int}. The form of $\bar{U}_{\rm ext}(\bm{x},\bm{w})$ for the case of a 1D harmonic trap plus the gravitational potential of an arbitrarily placed point mass will be given in Appendix~\ref{U_ext}.

\subsection{Derivation of $\bar{L}^{(3D)}_{1}$}
\label{L1_3D_derivation}

The $\bar{L}_{1}$ term of the Lagrangian has the form
\begin{equation}
\bar{L}^{(3D)}_{1} = 
\int\,d^{3}\bar{r}\,
{\rm Im}\left\{
\Psi^{\ast}({\bf\bar{r}},\bar{t})
\Psi_{\bar{t}}({\bf\bar{r}},\bar{t})
\right\}.
\label{L1_calc}
\end{equation}
The trial wave function and its time derivative are
\begin{eqnarray}
\Psi({\bf \bar{r}},\bar{t}) 
&=& 
\frac{1}{\sqrt{N_{c}}}\sum_{j=1}^{N_{c}}
A_{j}(\bar{t})
e^{f_{j}({\bf\bar{r}},\bar{t}) + {\bf\bar{k}}_{j}\cdot{\bf\bar{r}}},\nonumber\\
\Psi_{\bar{t}}({\bf \bar{r}},\bar{t}) 
&=& 
\frac{1}{\sqrt{N_{c}}}\sum_{j=1}^{N_{c}}
\left(
\dot{A}_{j}a(\bar{t}) +
A_{j}(\bar{t})\dot{f}_{j}({\bf\bar{r}},\bar{t})
\right)\nonumber\\
&\times&
e^{f_{j}({\bf\bar{r}},\bar{t}) + {\bf\bar{k}}_{j}\cdot{\bf\bar{r}}}
\end{eqnarray}
where the dot denotes partial differentiation with respect to $\bar{t}$.  Inserting these into Eq.\ (\ref{L1_calc}), neglecting the rapidly oscillating terms and then taking the imaginary part yields the following result
\begin{eqnarray}
\bar{L}^{(3D)}_{1} 
&=& 
\frac{1}{N_{c}}\sum_{j=1}^{N_{c}}
A_{j}^{2}(\bar{t})
\sum_{\eta=x,y,z}
\int\,d^{3}\bar{r}
\left(\dot{\bar{\alpha}}_{j\eta}\bar{\eta} + 
\dot{\bar{\beta}}_{j\eta}\bar{\eta}^{2}\right)\nonumber\\
&\times&
\prod_{\eta^{\prime}=x,y,z}
\exp\left\{
-\frac{\left(\bar{\eta^{\prime}}-
\bar{\eta^{\prime}}_{j}\right)^{2}}{\bar{w}_{j\eta^{\prime}}^{2}}
\right\}\nonumber\\
&=& 
\frac{1}{N_{c}}\sum_{j=1}^{N_{c}}
\sum_{\eta=x,y,z}
\left(\frac{1}{\pi^{1/2}\bar{w}_{j\eta}}\right)\\
&\times&
\int\,d^{3}\bar{r}
\left(\dot{\bar{\alpha}}_{j\eta}\bar{\eta} + 
\dot{\bar{\beta}}_{j\eta}\bar{\eta}^{2}\right)
\exp\left\{
-\frac{\left(\bar{\eta}-
\bar{\eta}_{j}\right)^{2}}{\bar{w}_{j\eta}^{2}}
\right\},\nonumber
\end{eqnarray}
where in the second line we have used Eqs.\ (\ref{3d_cloud_norm}) to eliminate the $A_{j}^{2}(\bar{t})$ factors.  The Gaussian integrals appearing in this last expression can easily be evaluated.  The final result for $\bar{L}^{(3D)}_{1}$ is
$\bar{L}_{1}$:
\begin{equation}
\bar{L}^{(3D)}_{1} = 
\frac{1}{N_{c}}\sum_{j=1}^{N_{c}}
\sum_{\eta=x,y,z}
\left(\dot{\bar{\alpha}}_{j\eta}\bar{\eta}_{j} + 
\dot{\bar{\beta}}_{j\eta}
\left(
\bar{\eta}_{j}^{2} +
\frac{1}{2}\bar{w}_{j\eta}^{2}
\right)
\right)
\label{L1_3D}
\end{equation}
Next we derive $\bar{L}^{(3D)}_{2}$.

\subsection{Derivation of $\bar{L}^{(3D)}_{2}$}
\label{L2_3D_derivation}
The expression for $\bar{L}^{(3D)}_{2}$ is given by
\begin{equation}
\bar{L}^{(3D)}_{2} \equiv
\sum_{\eta=x,y,z}
\int\,d^{3}\bar{r}
\Psi_{\bar{\eta}}^{\ast}
\Psi_{\bar{\eta}}.
\label{L2_eq}
\end{equation}
If we differentiate $\Psi$ with respect to $\bar{\eta}$ ($\eta$ is $x$, $y$, or $z$), multiply the result with its complex conjugate, and neglect the rapidly oscillating terms (which integrate to zero), we obtain
\begin{eqnarray}
\Psi_{\bar{\eta}}^{\ast}\Psi_{\bar{\eta}} 
&\approx&
\frac{1}{N_{c}}
\sum_{j=1}^{N_{c}}
A_{j}^{2}(\bar{t})
\exp
\left\{
-\sum_{\eta^{\prime}=x,y,z}
\frac
{\left(\eta^{\prime}-\eta^{\prime}_{j}\right)^{2}}
{\bar{w}_{j\eta^{\prime}}^{2}}
\right\}\nonumber\\
&\times&
\left(
\left(
\bar{\alpha}_{j\eta}+\bar{k}_{j\eta}+2\bar{\beta}_{j\eta}\eta
\right)^{2} +
\frac{\left(\eta-\eta_{j}\right)^{2}}{\bar{w}_{j\eta}^{4}}
\right).
\end{eqnarray}
Inserting this into Eq.\ (\ref{L2_eq}) gives
\begin{eqnarray}
\bar{L}^{(3D)}_{2} 
&=&
\frac{1}{N_{c}}
\sum_{\substack{j=1\\\eta=x,y,z}}^{N_{c}}
A_{j}^{2}(\bar{t})
\int\,d^{3}\bar{r}
\Bigg(
\left(
\bar{\alpha}_{j\eta}+\bar{k}_{j\eta}+2\bar{\beta}_{j\eta}\eta
\right)^{2}\nonumber\\
&+&
\frac{\left(\eta-\eta_{j}\right)^{2}}{\bar{w}_{j\eta}^{4}}
\Bigg)
\exp
\left\{
-\sum_{\eta^{\prime}=x,y,z}
\frac
{\left(\eta^{\prime}-\eta^{\prime}_{j}\right)^{2}}
{\bar{w}_{j\eta^{\prime}}^{2}}
\right\}
\end{eqnarray}
Again we are left with Gaussian integrals that can be straightforwardly evaluated.  The final result is
\begin{eqnarray}
\bar{L}^{(3D)}_{2}
&=&
\frac{1}{N_{c}}
\sum_{j=1}^{N_{c}}
\sum_{\eta=x,y,z}
\Bigg(
\frac{1}{2\bar{w}_{j\eta}^{2}} + 
2\bar{\beta}_{j\eta}^{2}\bar{w}_{j\eta}^{2}\nonumber\\
&+&
\Big(
\bar{\alpha}_{j\eta} +
\bar{k}_{j\eta}+
2\bar{\beta}_{j\eta}\bar{\eta}_{j}
\Big)^{2}
\Bigg).
\label{L2_3D}
\end{eqnarray}
In the above we have, again, used Eqs.\ (\ref{3d_cloud_norm}) to eliminate the $A_{j}^{2}(\bar{t})$ factors.

\subsection{3D Variational Potentials}
\label{3D_variational_potentials}

Next we consider the terms $L^{(3D)}_{3}$ and $L^{(3D)}_{4}$ which are
\begin{eqnarray}
\bar{L}^{(3D)}_{3} 
&\equiv&
\int\,d^{3}\bar{r}
\bar{V}_{\rm ext}({\bf\bar{r}},\bar{t})
\left|\Psi({\bf\bar{r}},\bar{t})\right|^{2}\\
\bar{L}^{(3D)}_{4} 
&\equiv& 
\frac{1}{2}\bar{g}N
\int\,d^{3}\bar{r}
\left|\Psi({\bf\bar{r}},\bar{t})\right|^{4}.
\end{eqnarray}
Note that, in terms of the trial wave function, these depend on $\left|\Psi({\bf\bar{r}},\bar{t})\right|^{2}$ or its square.  If we form $\Psi^{\ast}\Psi$ with our trial wave function and neglect the rapidly oscillating terms, we obtain
\begin{equation}
\left|\Psi\right|^{2}
\approx
\frac{1}{N_{c}}
\sum_{j_{1}=1}^{N_{c}}
A_{j_{1}}^{2}(\bar{t})
\exp
\left\{-\sum_{\eta=x,y,z}
\frac
{\left(\bar{\eta}-\bar{\eta}_{j_{1}}\right)^{2}}
{\bar{w}_{j_{1}^{2}}}
\right\}.
\end{equation}
It is clear that $|\Psi|^{2}$ only depends on the centers, $\bm{x}$, and widths, $\bm{w}$, if rapidly oscillating terms can be neglected.  This is evidently also the case for $|\Psi|^{4}$ which is the square of $|\Psi|^{2}$.  Thus, clearly, we have $\bar{L}^{(3D)}_{3} = \bar{L}^{(3D)}_{3}(\bm{x},\bm{w})$ and $\bar{L}^{(3D)}_{4} = \bar{L}^{(3D)}_{4}(\bm{x},\bm{w})$, that is, these quantities only depend on the Gaussian center and width variational parameters.

Thus we introduce here the following variational potentials.  First we define the ``external'' variational potential that accounts for the effects of $V_{\rm ext}$:
\begin{equation}
\bar{U}^{(3D)}_{\rm ext}(\bm{x},\bm{w}) \equiv 
2N_{c}\bar{L}^{(3D)}_{3}(\bm{x},\bm{w}).
\label{Uext_3d_app}
\end{equation}
We further introduce the ``interaction'' variational potential that accounts for atom--atom scattering interactions:
\begin{equation}
\bar{U}^{(3D)}_{\rm int}(\bm{x},\bm{w}) \equiv 
2N_{c}\bar{L}^{(3D)}_{4}(\bm{x},\bm{w}).
\label{Uint_3d_app}
\end{equation}
Note that $\bar{U}^{(3D)}_{\rm ext}(\bm{x},\bm{w})$ depends on the particular form of $V_{\rm ext}({\bf r},t)$.  The expression for $\bar{U}^{(3D)}_{\rm int}(\bm{x},\bm{w})$ is always the same.  The derivation of $\bar{U}^{(3D)}_{\rm int}(\bm{x},\bm{w})$ is more complex than that of the other Lagrangian terms and is given in Appendix\,\ref{U_int}.  It is only necessary to know that these potentials depend on $\bm{x}$ and $\bm{w}$ in order to derive the equations of motion for the Gaussian centers and widths.  The final equations of motion will be written in terms of the gradients of $\bar{U}^{(3D)}_{\rm ext}(\bm{x},\bm{w})$ and $\bar{U}^{(3D)}_{\rm int}(\bm{x},\bm{w})$.

\subsection{Derivation of $\bar{L}^{(3D)}_{5}$}
\label{L5_3D_derivation}

The term $\bar{L}^{(3D)}_{5}$ accounts for the possibility that the system is in a rotating frame.  Here we will assume that there is rotation only about the $z$ axis.  The $\bar{L}^{(3D)}_{5}$ term of the Lagrangian then has the form
\begin{equation}
\bar{L}^{(3D)}_{5} =
i\bar{\Omega}_{z}
\int\,d^{3}\bar{r}\,
\Psi
\left(
\bar{y}\Psi_{x}^{\ast} -
\bar{x}\Psi_{y}^{\ast}
\right).
\label{L5_eq}
\end{equation}
This term can be straightforwardly evaluated by inserting the expressions for the trial wave function and its space derivative into Eq.\ (\ref{L5_eq}), neglecting the oscillating terms, evaluating the resulting Gaussian integrals, and eliminating the $A_{j}^{2}(\bar{t})$ factors.  The result is
\begin{eqnarray}
\bar{L}^{(3D)}_{5} 
&=& 
\frac{\bar{\Omega}_{z}}{N_{c}}
\sum_{j=1}^{N_{c}}
\Bigg[
\bar{y}_{j}
\left(
\bar{\alpha}_{jx} + 
\bar{k}_{jx} +
2\bar{\beta}_{jx}\bar{x}_{j}
\right)\nonumber\\
&-& 
\bar{x}_{j}
\left(
\bar{\alpha}_{jy} + 
\bar{k}_{jy} +
2\bar{\beta}_{jy}\bar{y}_{j}
\right)
\bigg]
\label{L5_3D}
\end{eqnarray}

\subsection{The 3D Lagrangian}
\label{full_3d_lag}

Combining Eqs.\ (\ref{L1_3D}), (\ref{L2_3D}), (\ref{Uext_3d}), (\ref{Uint_3d}), and (\ref{L5_3D}) we have the full 3D Lagrangian
\begin{eqnarray}
\bar{L}^{(3D)} 
&=&
\frac{1}{N_{c}}\sum_{j=1}^{N_{c}}
\sum_{\eta=x,y,z}
\Bigg[\dot{\bar{\alpha}}_{j\eta^{\prime}}\bar{\eta}_{j} + 
\dot{\bar{\beta}}_{j\eta}
\left(
\bar{\eta}^{2}_{j} +
\frac{1}{2}\bar{w}_{j\eta}^{2}
\right)\nonumber\\
&+&
\frac{1}{2\bar{w}_{j\eta}^{2}} + 
2\bar{\beta}_{j\eta}^{2}
\bar{w}_{j\eta}^{2} +
\Big(
2\bar{\beta}_{j\eta}\bar{\eta}_{j} +
\bar{\alpha}_{j\eta} +
\bar{k}_{j\eta}
\Big)^{2}
\Bigg]\nonumber\\ 
&+&
\frac{\bar{\Omega}_{z}}{N_{c}}
\sum_{j=1}^{N_{c}}
\Bigg[
\bar{y}_{j}
\left(
\bar{\alpha}_{jx} + 
\bar{k}_{jx} +
2\bar{\beta}_{jx}\bar{x}_{j}
\right) -
\bar{x}_{j}
\big(
\bar{\alpha}_{jy}\nonumber\\ 
&+& 
\bar{k}_{jy} +
2\bar{\beta}_{jy}\bar{y}_{j}
\big)
\Bigg] +
\frac{1}{2N_{c}}
\Big(
\bar{U}^{(3D)}_{\rm ext}({\bf x},{\bf w})\nonumber\\
&+& 
\bar{U}^{(3D)}_{\rm int}({\bf x},{\bf w})
\Big)
\label{L_final_3d_app}
\end{eqnarray}
We will use this form of the Lagrangian to derive the 3D equations of motion.

\section{Derivation of $\bar{U}_{int}^{(3D)}({\bf x},{\bf w})$ and $\bar{U}_{int}^{(1D)}({\bf x},{\bf w})$}
\label{U_int}

Here we present the derivation of the expressions for $\bar{U}^{(3D)}_{\rm int}({\bf x},{\bf w})$ and $\bar{U}^{(1D)}_{\rm int}({\bf x},{\bf w})$.  The expression for $\bar{U}^{(3D)}_{\rm int}({\bf x},{\bf w})$ is
\begin{equation}
\bar{U}^{(3D)}_{\rm int}({\bf x},{\bf w}) \equiv 
2N_{c}
\left(
\frac{1}{2}\bar{g}N
\int\,d^{3}\bar{r}
\left|\Psi\right|^{4}
\right)
\label{u_int_app_a}
\end{equation}
In order to perform this integral we must first calculate $|\Psi|^{4}$.  We can write this quantity as the square of $|\Psi|^{2}$:
\begin{eqnarray}
\left|\Psi\right|^{4} 
&=&
\frac{1}{N_{c}^{2}}
\Bigg[
\sum_{j_{1}=1}^{N_{c}}
A_{j_{1}}^{2}
e^{2\sum_{\eta=x,y,z}{\rm Re}\{f_{j_{1}\eta}\}} + 
\sum_{\substack{j_{1},j_{2}=1\\j_{1}\ne j_{2}}}^{N_{c}}
A_{j_{1}}A_{j_{2}}\nonumber\\
&\times&
\exp
\Bigg\{
\sum_{\eta=x,y,z}
\Big(
f_{j_{1}\eta}^{\ast}+
f_{j_{2}\eta} +
i\left(\bar{k}_{j_{2}\eta}-\bar{k}_{j_{1}\eta}\right)\bar{\eta}
\Big)
\Bigg\}
\Bigg]^{2}\nonumber
\end{eqnarray}
where the have written the $|\Psi|^{2}$ appearing inside the large square brackets as a sum of two terms.  The first term is a sum over terms that definitely do not oscillate in space ($j_{1}=j_{2}$) and the second term, where $j_{1}\ne j_{2}$, all of the terms definitely do oscillate in space.

When this expression is expanded by writing the terms of the overall square only the square of the first term and the square of the last term need be retained.  We only need to integrate terms that definitely do not oscillate as the oscillatory terms will integrate to zero.  Thus we can rewrite the above expression as follows:
\begin{eqnarray}
|\Psi|^{4} 
&\approx&
\frac{1}{N_{c}^{2}}
\Bigg[
\sum_{j_{1},j_{2}=1}^{N_{c}}
A_{j_{1}}^{2}A_{j_{2}}^{2}\nonumber\\
&\times&
\exp\left\{
\sum_{\eta=x,y,z}
\Big(
f_{j_{1}\eta}^{\ast} + f_{j_{1}\eta} +
f_{j_{2}\eta}^{\ast} + f_{j_{2}\eta}
\Big)
\right\}\nonumber\\
&+&
\sum_{\substack{j_{1},j_{2}=1\\j_{1}\ne j_{2}}}
\sum_{\substack{j_{1}^{\prime},j_{2}^{\prime}=1\\
j_{1}^{\prime}\ne j_{2}^{\prime}}}
A_{j_{1}}A_{j_{2}}A_{j_{1}^{\prime}}A_{j_{2}^{\prime}}\nonumber\\
&\times&
\exp\Bigg\{
\sum_{\eta=x,y,z}
\Big(
f_{j_{1}\eta}^{\ast}+
f_{j_{2}\eta}+
f_{j_{1}^{\prime}\eta}^{\ast}+
f_{j_{2}^{\prime}\eta}\nonumber\\
&+&
i\left(
\bar{k}_{j_{2}\eta}-\bar{k}_{j_{1}\eta}+
\bar{k}_{j_{2}^{\prime}\eta}-\bar{k}_{j_{1}^{\prime}\eta}
\right)\bar{\eta}
\Big)
\Bigg\}
\Bigg]\nonumber
\end{eqnarray}
The second term above still has oscillating terms.  We only want to keep the non--oscillating terms.  The terms that don't oscillate are those where $j_{1}^{\prime}=j_{2}$ and $j_{2}^{\prime}=j_{1}$ (since $j_{1}=j_{2}$ and $j_{1}^{\prime}=j_{2}^{\prime}$ are excluded already).  Thus we can evaluate the primed sums keeping only those terms where $j_{1}^{\prime}=j_{2}$ and $j_{2}^{\prime}=j_{1}$:
\begin{eqnarray}
|\Psi|^{4} 
&\approx&
\frac{1}{N_{c}^{2}}
\Bigg[
\sum_{j_{1},j_{2}=1}^{N_{c}}
A_{j_{1}}^{2}A_{j_{2}}^{2}\nonumber\\
&\times&
\exp\left\{
\sum_{\eta=x,y,z}
\Big(
f_{j_{1}\eta}^{\ast} + f_{j_{1}\eta} +
f_{j_{2}\eta}^{\ast} + f_{j_{2}\eta}
\Big)
\right\}\nonumber\\
&+&
\sum_{\substack{j_{1},j_{2}=1\\j_{1}\ne j_{2}}}
A_{j_{1}}^{2}A_{j_{2}}^{2}\nonumber\\
&\times&
\exp\left\{
\sum_{\eta=x,y,z}
\Big(
f_{j_{1}\eta}^{\ast}+
f_{j_{2}\eta}+
f_{j_{2}\eta}^{\ast}+
f_{j_{1}\eta}
\Big)
\right\}
\Bigg]\nonumber
\end{eqnarray}
The two double sums appearing in the expression above are the same with one exception: the second double sum requires $j_{1}\ne j_{2}$ while the first double sum has no such restriction. We can write the first sum as two terms: a single sum where $j_{1}=j_{2}$ and a double sum where $j_{1}\ne j_{2}$.  This second term will then be identical to the second term in the original expression.
Carrying out this procedure writing $|\Psi|^{4}$ in terms of coordinates yields
\begin{eqnarray}
|\Psi|^{4} 
&\approx&
\frac{1}{N_{c}^{2}}
\Bigg[
\sum_{j_{1}=1}^{N_{c}}
A_{j_{1}}^{4}
\exp\left\{
-\sum_{\eta=x,y,z}
\Big(
\frac{2\left(\bar{\eta}-\bar{\eta}_{j_{1}}\right)^{2}}
{\bar{w}_{j_{1}\eta}^{2}}
\Big)
\right\}\nonumber\\
&+&
2\sum_{\substack{j_{1},j_{2}=1\\j_{1}\ne j_{2}}}
A_{j_{1}}^{2}A_{j_{2}}^{2}\nonumber\\
&\times&
\exp\left\{
-\sum_{\eta=x,y,z}
\Big(
\frac{\left(\bar{\eta}-\bar{\eta}_{j_{1}}\right)^{2}}
{\bar{w}_{j_{1}\eta}^{2}}+
\frac{\left(\bar{\eta}-\bar{\eta}_{j_{2}}\right)^{2}}
{\bar{w}_{j_{2}\eta}^{2}}
\Big)
\right\}
\Bigg]\nonumber
\end{eqnarray}
This form for $|\Psi|^{4}$ can now be inserted into Eq.\ (\ref{u_int_app_a}) and then resulting Gaussian integrals can be straightforwardly evaluated.  The final result is
\begin{eqnarray}
\bar{U}_{\rm int}^{(3D)}
&=&
\frac{\bar{g}N}{(2\pi)^{3/2}N_{c}}
\Bigg[
\sum_{j_{1}=1}^{N_{c}}
\left(
\frac{1}{\bar{w}_{j_{1}x}\bar{w}_{j_{1}y}\bar{w}_{j_{1}z}}
\right)\nonumber\\ 
&+&
2^{5/2}\sum_{\substack{j_{1},j_{2}=1\\j_{1}\ne j_{2}}}^{N_{c}}
\prod_{\eta=x,y,z}
\left(
\frac
{
\exp
\left\{
-\frac
{\left(\bar{\eta}_{j_{1}}-\bar{\eta}_{j_{2}}\right)^{2}}
{\bar{w}_{j_{1}\eta}^{2}+\bar{w}_{j_{2}\eta}^{2}}
\right\}
}
{
\left(\bar{w}_{j_{1}\eta}^{2}+\bar{w}_{j_{2}\eta}^{2}\right)^{1/2}
}
\right)
\Bigg]\nonumber\\
\label{u_int_final_3d}
\end{eqnarray}
Using a similar derivation, we obtain the 1D version of $\bar{U}_{\rm int}$:
\begin{eqnarray}
\bar{U}_{\rm int}^{(1D)} 
&=& 
\frac{\left(\bar{g}N\right)}{(2\pi)^{1/2}N_{c}}
\Bigg[
\sum_{j_{1}=1}^{N_{c}}
\frac{1}{\bar{w}_{j_{1}}}\nonumber\\
&+&
2^{3/2}\sum_{\substack{j_{1},j_{2}=1\\j_{1}\ne j_{2}}}^{N_{c}}
\frac
{
\exp
\left\{
-\frac
{\left(\bar{x}_{j_{1}}-\bar{x}_{j_{2}}\right)^{2}}
{\bar{w}_{j_{1}}^{2}+\bar{w}_{j_{2}}^{2}}
\right\}
}
{
\left(\bar{w}_{j_{1}}^{2}+\bar{w}_{j_{2}}^{2}\right)^{1/2}
}
\Bigg]
\label{u_int_final_1d}
\end{eqnarray}
The derivatives of these expressions with respect to the center and width variational parameters are computed straightforwardly.

\section{The LVM Equations of Motion}
\label{eoms_app}

In this section we derive the LVM equations of motion for the variational parameters appearing in the 3D and 1D trial wave functions given above.  The major steps in this undertaking are the following.  First we write down the equation of motion for each variational parameter, $\bar{q}$, using the standard Euler--Lagrange (EL) equation (Eq.\ (\ref{standard_EL})).  We will show that further manipulations of these equations can produce a closed set of second--order differential equations in time for the center position and width coordinates.  Furthermore we will show that all of the other variational parameters can be expressed in terms of the position and width coordinates and their time derivatives.  We begin with the 3D case.

\subsection{3D LVM equations of motion}
\label{3d_eoms}

\subsubsection{3D cloud center EOMs}
\label{3d_center_eoms}

To obtain the equations of motion for the cloud--center coordinates we need the EOMs associated with the $\bar{\alpha}_{j\eta}$ and the $\bar{\eta}_{j}$. We begin with the EOMs associated with the $\bar{\alpha}_{j\eta}$.  The Euler--Lagrange equations for these are
\begin{equation}
\frac{d}{d\bar{t}}
\left(
\frac
{\partial\bar{L}^{(3D)}}
{\partial\dot{\bar{\alpha}}_{j\eta}}
\right) - 
\frac
{
\partial\bar{L}^{(3D)}
}
{
\partial \bar{\alpha}_{j\eta}
} = 0,
\quad
j = 1,\dots,N_{c},
\ \eta = x,y,z.
\end{equation}
We can compute the derivatives of $\bar{L}$ using Eq.\ (\ref{L_final_3d}).  The straightforward result is
\begin{eqnarray}
\dot{x}_{j} 
&=& 
2\left(
2\bar{\beta}_{jx}\bar{x}_{j} + \bar{\alpha}_{jx} + \bar{k}_{jx} 
\right) +
\bar{\Omega}_{z}\bar{y}_{j},
\label{alphajx_3d}\\
\dot{y}_{j} 
&=& 
2\left(
2\bar{\beta}_{jy}\bar{y}_{j} + \bar{\alpha}_{jy} + \bar{k}_{jy} 
\right) -
\bar{\Omega}_{z}\bar{x}_{j},
\label{alphajy_3d}\\
\dot{z}_{j} 
&=& 
2\left(
2\bar{\beta}_{jz}\bar{z}_{j} + \bar{\alpha}_{jz} + \bar{k}_{jz}
\right),
\quad
j = 1,\dots,N_{c}.
\label{alphajz_3d}
\end{eqnarray}

Next we obtain the EL equations associated with $\bar{x}_{j}$, $\bar{y}_{j}$, and $\bar{z}_{j}$.  The $\bar{\eta}_{j}$ Euler--Lagrange EOMs can be written as
\begin{eqnarray}
\frac{\partial\bar{L}}{\partial \bar{\eta}_{j}} 
&=&
\frac{d}{d\bar{t}}
\left(
\frac
{\partial\bar{L}}
{\partial\dot{\bar{\eta}}_{j}}
\right),\nonumber\\
\frac{\partial\bar{L}}{\partial \bar{\eta}_{j}} 
&=& 0,
\quad
j = 1,\dots,N_{c}
\ \eta = x,y,z
\end{eqnarray}
where the second equality holds as there are no $\dot{\bar{\eta}}$ terms in the Lagrangian.  Taking the necessary derivatives we obtain the EL equations associated with $\bar{j}$, $\bar{y}_{j}$, and $\bar{z}_{j}$, respectively
\begin{eqnarray}
0 &=&
\Big[
\dot{\alpha}_{jx} + 
2\dot{\bar{\beta}}_{jx}\bar{x}_{j} +
2
\left(
2\bar{\beta}_{jx}\bar{x}_{j} +
\bar{\alpha}_{jx} + 
\bar{k}_{jx}
\right)
\left(2\bar{\beta}_{jx}\right)\nonumber\\ 
&+&
2\bar{\Omega}_{z}\bar{\beta}_{jx}\bar{y}_{j} 
\Big] +
\frac{1}{2}
\left(
\frac{\partial\bar{U}^{(3D)}_{\rm ext}}{\partial\bar{x}_{j}} +
\frac{\partial\bar{U}^{(3D)}_{\rm int}}{\partial\bar{x}_{j}}
\right)\nonumber\\ 
&-& 
\bar{\Omega}_{z}
\left(
2\bar{\beta}_{jy}\bar{y}_{j} +
\bar{\alpha}_{jy} +
\bar{k}_{jy}
\right)
\label{ELxj_3d}\\
0
&=&
\Big[
\dot{\alpha}_{jy} + 
2\dot{\bar{\beta}}_{jy}\bar{y}_{j} +
2
\left(
2\bar{\beta}_{jy}\bar{y}_{j} +
\bar{\alpha}_{jy} + 
\bar{k}_{jy}
\right)
\left(2\bar{\beta}_{jy}\right)\nonumber\\
&-&
2\bar{\Omega}_{z}\bar{\beta}_{jy}\bar{x}_{j} 
\Big] +
\frac{1}{2}
\left(
\frac{\partial\bar{U}^{(3D)}_{\rm ext}}{\partial\bar{y}_{j}} +
\frac{\partial\bar{U}^{(3D)}_{\rm int}}{\partial\bar{y}_{j}}
\right)\nonumber\\ 
&+& 
\bar{\Omega}_{z}
\left(
2\bar{\beta}_{jx}\bar{x}_{j} +
\bar{\alpha}_{jx} +
\bar{k}_{jx}
\right)
\label{ELyj_3d}\\
0
&=&
\Big[
\dot{\alpha}_{jz} + 
2\dot{\bar{\beta}}_{jz}\bar{z}_{j} +
2
\left(
2\bar{\beta}_{jz}\bar{z}_{j} +
\bar{\alpha}_{jz} + 
\bar{k}_{jz}
\right)
\left(2\bar{\beta}_{jz}\right)
\Big]\nonumber\\ 
&+&
\frac{1}{2}
\left(
\frac{\partial\bar{U}^{(3D)}_{\rm ext}}{\partial\bar{z}_{j}} +
\frac{\partial\bar{U}^{(3D)}_{\rm int}}{\partial\bar{z}_{j}}
\right)
\quad
j=1,\dots,N_{c}
\label{ELzj_3d}
\end{eqnarray}
We have written the above equations in a form for convenient use in deriving the final equations of motion. Equations (\ref{alphajx_3d}), (\ref{alphajy_3d}), and (\ref{alphajz_3d}) can be used along with Eqs.\ (\ref{ELxj_3d}), (\ref{ELyj_3d}), and (\ref{ELzj_3d}) to derive second--order differential equations that involve only the center and width coordinates.

We illustrate how this can be done by deriving the equation for $\ddot{\bar{x}}_{j}$.  If we differentiate both sides of the equation for $\dot{\bar{x}}_{j}$ (Eq.\ (\ref{alphajx_3d})) with respect to time, the resulting second--order equation will contain both $\ddot{\bar{x}}_{j}$ and $\dot{\bar{x}}_{j}$.  Using Eq.\ (\ref{alphajx_3d}) again to eliminate $\dot{\bar{x}}_{j}$ from this second--order equation gives the following result. 
\begin{eqnarray}
\ddot{\bar{x}}_{j} 
&=&
2\Big[
\dot{\bar{\alpha}}_{jx} +
2\dot{\bar{\beta}}_{jx}\bar{x}_{j} +
2\left(
2\bar{\beta}_{jx}\bar{x}_{j} +
\bar{\alpha}_{jx} + 
\bar{k}_{jx}
\right)
\left(
2\bar{\beta}_{jx}
\right)\nonumber\\
&+&
2\bar{\Omega}_{z}\bar{\beta}_{jx}\bar{y}_{j}
\Big] +
\bar{\Omega}_{z}\dot{\bar{y}}_{j}\nonumber\\
&=&
-\left(
\frac{\partial\bar{U}^{(3D)}_{\rm ext}}{\partial\bar{x}_{j}} +
\frac{\partial\bar{U}^{(3D)}_{\rm int}}{\partial\bar{x}_{j}}
\right)\nonumber\\
&+&
2\bar{\Omega}_{z}
\left(
2\bar{\beta}_{jy}\bar{y}_{j} +
\bar{\alpha}_{jy} +
\bar{k}_{jy}
\right) +
\bar{\Omega}_{z}\dot{\bar{y}}_{j}
\label{tempx1}
\end{eqnarray}
where we have obtained the second line by noting that the quantity in square brackets appearing in the first line is identical to the quantity in square brackets in Eq.\ (\ref{ELxj_3d}).  The second equation can be further simplified by noting that the quantity in the parenthesis appearing in the second term is also present in Eq.\ (\ref{alphajx_3d}).  The second--order equation of motion is thus written in the compact final form
\begin{eqnarray}
\ddot{\bar{x}}_{j} 
&=&
2\bar{\Omega}_{z}\dot{\bar{y}}_{j} +
\bar{\Omega}_{z}^{2}\bar{x}_{j} -
\frac{\partial\bar{U}^{(3D)}}{\partial\bar{x}_{j}},
\quad{\rm where}\nonumber\\
\bar{U}^{(3D)}({\bf x},{\bf w})
&\equiv&
\bar{U}^{(3D)}_{\rm ext}({\bf x},{\bf w}) + 
\bar{U}^{(3D)}_{\rm int}({\bf x},{\bf w}).
\label{xddot_final}
\end{eqnarray}
The corresponding equations for the other cloud--center coordinates can be derived similarly.

The resulting second--order equations of motion for the center coordinates of cloud $j$ are as follows. 
\begin{subequations}
\begin{align}
\label{xj_eom}
\ddot{\bar{x}}_{j} &=
2\bar{\Omega}_{z}\dot{\bar{y}}_{j} +
\bar{\Omega}_{z}^{2}\bar{x}_{j} -
\frac{\partial\bar{U}^{(3D)}}{\partial\bar{x}_{j}},\\
\label{yj_eom}
\ddot{\bar{y}}_{j} &=
-2\bar{\Omega}_{z}\dot{\bar{x}}_{j} +
\bar{\Omega}_{z}^{2}\bar{y}_{j} -
\frac{\partial\bar{U}^{(3D)}}{\partial\bar{y}_{j}},\\
\label{zj_eom}
\ddot{\bar{z}}_{j} &= 
-\frac{\partial\bar{U}^{(3D)}}{\partial\bar{z}_{j}},
\quad
j = 1,\dots,N_{c}.
\end{align}
\end{subequations}
Note that these equations depend only on ${\bf x}$, ${\bf w}$ and their time derivatives. We now turn to the equations for the cloud widths.

\subsubsection{3D cloud width EOMs}
\label{3d_width_eoms}

It is possible to derive a set of second--order equations of motion for the Gaussian cloud widths similar to that for the cloud centers.  As will be seen below, the center equations and width equations form a closed system.  All of the other variational parameters can be expressed in terms of the centers and widths and their time derivatives.

To obtain the width equations we start with the EL equation for the $\bar{\beta}_{j\eta}$. As before, we will only derive the equation of motion for $\bar{w}_{jx}$ to illustrate how the derivation is carried out.  The EL equation for $\bar{\beta}_{jx}$ is
\begin{equation}
\frac{d}{d\bar{t}}
\left(
\frac{\partial\bar{L}}{\partial\dot{\bar{\beta}}_{jx}}
\right) - 
\frac{\partial\bar{L}}{\partial \bar{\beta}_{jx}} = 0,
\quad
j = 1,\dots,N_{c}.
\end{equation}
Differentiating the 3D Lagrangian in Eq.\ (\ref{L_final_3d}) and inserting into the above equations gives
\begin{eqnarray}
2\bar{x}_{j}\dot{\bar{x}}_{j} +
\bar{w}_{jx}\dot{\bar{w}}_{jx} 
&=& 
\left(
4\bar{x}_{j}
\right)
\left(
2\bar{\beta}_{jx}\bar{x}_{j} +
\bar{\alpha}_{jx} +
\bar{k}_{jx}
\right)\nonumber\\ 
&+&
2\bar{\Omega}_{z}\bar{y}_{j}\bar{x}_{j} +
4\bar{\beta}_{jx}\bar{w}_{jx}^{2}.
\label{ELbetajx}
\end{eqnarray}
If we replace $\dot{\bar{x}}_{j}$ on the left--hand--side with its expression given in Eq.\ (\ref{alphajx_3d}) we obtain the following remarkably simple result
\begin{equation}
\dot{\bar{w}}_{jx} = 4\bar{\beta}_{jx}\bar{w}_{jx}.
\label{ELbetajx_final}
\end{equation}
This result holds for $\bar{y}$ and $\bar{z}$ as well and enables us to express the $\bar{\beta}_{j\eta}$ in terms of the widths.

To proceed we turn to the Euler--Lagrange equation for $\bar{w}_{jx}$ which reads
\begin{equation}
\frac{d}{d\bar{t}}
\left(
\frac
{\partial\bar{L}}
{\partial\dot{\bar{w}}_{jx}}
\right) - 
\frac
{
\partial\bar{L}
}
{
\partial \bar{w}_{jx}
} = 0 = 
\frac
{
\partial\bar{L}
}
{
\partial \bar{w}_{jx}
},
\end{equation}
where the last equality holds because no terms containing $\dot{\bar{w}}_{jx}$ appear in the Lagrangian given in equation Eq.\ (\ref{L_final_3d}).  Again taking derivatives of $\bar{L}^{(3D)}$ we obtain
\begin{equation}
4\dot{\bar{\beta}}_{jx}\bar{w}_{jx} +
16\bar{\beta}_{jx}^{2}\bar{w}_{jx} =
\frac{4}{\bar{w}_{jx}^{3}} -
2\frac{\partial\bar{U}^{(3D)}}
{\partial\bar{w}_{jx}}.
\label{wjx_eom}
\end{equation}
Now note that, if we differentiate both sides of Eq.\ (\ref{ELbetajx_final}) with respect to time we obtain
\begin{eqnarray}
\ddot{\bar{w}}_{jx} 
&=& 
4\dot{\bar{\beta}}_{jx}\bar{w}_{jx} +
4\bar{\beta}_{jx}\dot{\bar{w}}_{jx}\nonumber\\
&=&
4\dot{\bar{\beta}}_{jx}\bar{w}_{jx} +
16\bar{\beta}_{jx}^{2}\bar{w}_{jx},
\end{eqnarray}
where the second equality comes from reusing Eq.\ (\ref{ELbetajx_final}) to replace $\dot{\bar{w}}_{jx}$ in the first line. Finally note that the right--hand--side of this last expression is identical to the left--hand--side of Eq.\ (\ref{wjx_eom}) and thus we obtain the second--order equation for $\bar{w}_{jx}$:
\begin{equation}
\ddot{\bar{w}}_{jx} =
\frac{4}{\bar{w}_{jx}^{3}} - 
2\frac{\partial\bar{U}^{(3D)}}
{\partial\bar{w}_{jx}}.
\label{width_x_eq}
\end{equation}
The derivation of equations for $\bar{y}$ and $\bar{z}$ are similar.  All three equations can be written in the following compact form:
\begin{equation}
\ddot{\bar{w}}_{j\eta} =
\frac{4}{\bar{w}_{j\eta}^{3}} - 
2\frac{\partial\bar{U}^{(3D)}}
{\partial\bar{w}_{j\eta}},
\quad
\eta=x,y,z
\quad
j=1,\dots,N_{c}.
\label{width_eqs_3d}
\end{equation}
We can add these to EOMs for the cloud centers to get the full set of equations of motion in 3D. They consist of a pair of second--order ordinary differential equation for the cloud centers and widths as well as expressions for the $\bar{\beta}_{j\eta}$ and the $\bar{\alpha}_{j\eta}$ in terms of the centers, widths and their first derivatives:
\begin{subequations}
\begin{align}
\label{xj_eom_final_app}
\ddot{\bar{x}}_{j} &=
2\bar{\Omega}_{z}\dot{\bar{y}}_{j} +
\bar{\Omega}_{z}^{2}\bar{x}_{j} -
\frac{\partial\bar{U}^{(3D)}}{\partial\bar{x}_{j}},\\
\label{yj_eom_final_app}
\ddot{\bar{y}}_{j} &=
-2\bar{\Omega}_{z}\dot{\bar{x}}_{j} +
\bar{\Omega}_{z}^{2}\bar{y}_{j} -
\frac{\partial\bar{U}^{(3D)}}{\partial\bar{y}_{j}},\\
\label{zj_eom_final_app}
\ddot{\bar{z}}_{j} &= 
-\frac{\partial\bar{U}^{(3D)}}{\partial\bar{z}_{j}},\\
\label{wjeta_eoms_final_app}
\ddot{\bar{w}}_{j\eta} &=
\frac{4}{\bar{w}_{j\eta}^{3}} - 
2\frac{\partial\bar{U}^{(3D)}}
{\partial\bar{w}_{j\eta}},\\
\label{betajeta_eoms_final_app}
\bar{\beta}_{j\eta} &=
\frac{\dot{\bar{w}}_{j\eta}}{4\bar{w}_{j\eta}},\\
\label{alphajx_eom_final_app}
\bar{\alpha}_{jx} &=
\tfrac{1}{2}(\dot{\bar{x}}_{j} - \bar{\Omega}_{z}\bar{y}_{j}) - 
2\bar{\beta}_{jx}\bar{x}_{j} - \bar{k}_{jx},\\
\label{alphajy_eom_final_app}
\bar{\alpha}_{jy} &=
\tfrac{1}{2}(\dot{\bar{y}}_{j} + \bar{\Omega}_{z}\bar{x}_{j}) - 
2\bar{\beta}_{jy}\bar{y}_{j} - \bar{k}_{jy},\\
\label{alphajz_eom_final_app}
\bar{\alpha}_{jz} &=
\tfrac{1}{2}\dot{\bar{z}}_{j} - 
2\bar{\beta}_{jx}\bar{x}_{j} - \bar{k}_{jx},\\
\eta &= x,y,z\quad j=1,\dots,N_{c}\nonumber
\end{align}
\end{subequations}
The equations for the cloud centers and cloud widths (Eqs.\ (\ref{xj_eom_final_app}), (\ref{yj_eom_final_app}), (\ref{zj_eom_final_app}), and (\ref{wjeta_eoms_final_app})) form a closed set that contain only the $\bar{\eta}_{j}$, $\dot{\bar{\eta}}_{j}$, $\bar{w}_{j\eta}$, and $\dot{\bar{w}}_{j\eta}$.  Once these quantities are obtained,
all of the other variational parameters can be calculated.

\section{Details of the GPE simulation}
\label{gpe_sim}

The conditions assumed in the simulation shown in Fig.\,\ref{lvm_gpe_cmp} include a condensate of 10,000 $^{87}$Rb atoms which were confined in a harmonic trap with a frequency of $\omega_{T}/2\pi = 1$\,Hz. The condensate pieces were given an initial velocity of $v = 2\times 10^{-3}$ m/s and the wait time after reaching the trap turning points is $T = 500$\,ms. The source mass is absent in the simulation presented.  

The GPE was solved numerically using the split--step, Crank--Nicolson algorithm~\cite{MURUGANANDAM20091888}.  Briefly, in this algorithm, the Hamiltonian is split as $\hat{H}=\hat{T} + \hat{V}$ where $\hat{T}$ is the kinetic energy and $\hat{V}$ is the potential plus the nonlinear interaction term.  The condensate wave function is advanced from time $t$ to time $t+\delta t$ by multiplying $\psi(x,t)$ point by $e^{-iV(x,t)\delta t/\hbar}$ and then the result is multiplied by $e^{-i\hat{T}\delta t/\hbar}$.  This second step is carried out using the Crank--Nicolson algorithm. The time--dependent variational equations of motion were solved by the Euler method.  

The comparison of the GPE and the LVM simulations for the initial split phase of the AI sequence is shown in Fig.\,\ref{lvm_gpe_cmp}.  The initial split itself divided into three parts. In the first part the condensate pieces fly out to the turning points in the presence of the harmonic potential whose frequency is $\omega_{T}/2\pi = 1$ Hz and finally stop after a quarter period (250 ms) as shown in Figs.\,\ref{bigG_proposal}(b)--(c).  Both the GPE and the LVM densities are plotted together on the same graph at four different times as seen in the left panel of Fig.\,\ref{lvm_gpe_cmp}.  Each density consists of two symmetrically placed peaks and the left peak of each density is labeled with its time stamp.  In the second part of the initial split sequence the two condensate pieces evolve at rest for a wait time of $T=500$ ms while the trap is off.  This evolution is shown in the middle panel of Fig.\,\ref{lvm_gpe_cmp}. Finally the condensate pieces recombine after the trap is turned back on.  This is shown in the right panel of Fig.\,\ref{lvm_gpe_cmp}.

\section{Derivation of $\bar{U}_{ext}^{(1D)}({\bf x},{\bf w})$ for an harmonic plus point--mass potential}
\label{U_ext}

In this appendix we derive the variational potentials for $\bar{U}_{ext}^{(1D)}({\bf x},{\bf w})$ for the case of an external potential consisting of an harmonic potential plus the gravatational potential of a point mass situated far from the condensate.  The 1D version is used to obtain the equations of motion whose solutions are compared with the full 1D GPE simulation of the illustrative AI measurement of $G$ in a microgravity environment presented in Section~\ref{bigG}.

The 1D external variational potential is given by Eq.\ (\ref{Uext_1d})
\begin{eqnarray}
\bar{U}^{(1D)}_{\rm ext}
&=&
\sum_{j=1}^{N_{c}}
\left(
\frac{2}{\pi^{1/2}\bar{w}_{j}}
\right)
\int_{-\infty}^{+\infty}\,d\bar{x}\,
e^{-\left(\bar{x}-\bar{x}_{j}\right)^{2}/\bar{w}_{j}^{2}}
\bar{V}_{\rm ext}(\bar{x},\bar{t}).\nonumber\\
\label{Uext_1d_appB}
\end{eqnarray}
The first step is to find an approximate expression for $\bar{V}_{\rm ext}(\bar{x},\bar{t})$ for the combination of an harmonic potential plus a point mass.

We assume that the harmonic trap is centered at the origin of coordinates and that a point mass with mass $M_{SM}$ is located at ${x}_{SM}$.  The exact potential can thus be written (in SI units) as
\begin{equation}
V_{\rm ext}(x) = 
\frac{1}{2}M\omega_{T,x}^{2}x^{2} -
\frac{GMM_{SM}}{\left|x_{SM}-x\right|}
\equiv V_{\rm H}(x) + V_{\rm G}(x).
\end{equation}
We want to approximate $V_{\rm G}({\bf r})$ by assuming that the distance of any cloud to the origin is much smaller than the distance of the source mass to the origin.  We have chosen the origin to be at the center of the harmonic potential confining the BEC.

First we consider only the gravitational part of the potential: 
\begin{equation}
V_{\rm G}(x) = 
-\frac{GMM_{SM}}{\left|x_{SM}-x\right|}.
\end{equation}
we can approximate this exact expression by making a Taylor expansion about $x=0$ to second order in $x/x_{SM}$
\begin{equation}
V_{\rm G}(x) \approx
-\frac{GMM_{SM}}{|x_{SM}|^{3}}
\left(
x_{SM}^{2} + x_{SM}x + x^{2}
\right)
\end{equation}

This expression is valid only for points $x$ such that $x_{SM} > |x|$ which we take to be the case since we are actually assuming $x_{SM} \gg |x|$. It will be convenient here to introduce the gravitational frequency
\begin{equation}
\omega_{SM} \equiv \left(\frac{GM_{SM}}{|x_{SM}|^{3}}\right)^{1/2}.
\end{equation}
Thus we can rewrite the approximate gravitational potential in terms of this quantity as follows.
\begin{eqnarray}
V_{\rm G}({\bf r}) 
&\approx&
-M\omega_{SM}^{2}
\left(
x_{SM}^{2} + x_{SM}x + x^{2}
\right)\nonumber\\
\bar{V}_{\rm G}({\bf r})
&\approx&
-\tfrac{1}{2}\bar{\omega}_{SM}^{2}
\left(
\bar{x}_{SM}^{2} + \bar{x}_{SM}\bar{x} + \bar{x}^{2}
\right),
\label{grav_freq}
\end{eqnarray}
where, in the second line, we have expressed the gravitational potential in scaled units. Now we are ready to write down the full external potential.  The full external potential in scaled units is thus
\begin{eqnarray}
\bar{V}_{\rm ext}(\bar{x}) 
&=&
\tfrac{1}{4}
\left(\bar{\omega}_{T}^{2} -
2\bar{\omega}_{SM}^{2}
\right)\bar{x}^{2} -
\tfrac{1}{2}\bar{\omega}_{SM}^{2}\bar{x}_{SM}\bar{x}\nonumber\\
&-&
\tfrac{1}{2}\bar{\omega}_{SM}^{2}\bar{x}_{SM}^{2}
\label{vext}
\end{eqnarray}
Inserting the above expression into Eq.\ (\ref{Uext_1d_appB}) yields the final expression for $\bar{U}^{(1D)}_{\rm ext}$ for the case of an external harmonic trap plus the gravitational potential produced by a point mass far away from the condensate is given by
\begin{eqnarray}
\bar{U}^{(1D)}_{\rm ext}\left({\bf x},{\bf w}\right) 
&=& 
\sum_{j=1}^{N_{c}}
\Big(
\tfrac{1}{2}
\left(
\bar{\omega}_{T}^{2} -
2\bar{\omega}_{SM}^{2}
\right)
\left(
\bar{x}_{j}^{2} +
\tfrac{1}{2}\bar{w}_{j}^{2}
\right)\nonumber\\
&-&
\bar{\omega}_{SM}^{2}\bar{x}_{SM}\bar{x}_{j}
\Big) -
N_{c}\bar{\omega}_{SM}^{2}\bar{x}_{SM}^{2}.
\label{u_ext_final_1d}
\end{eqnarray}
We note that the last term is constant and never appears in the equations of motion.

\end{document}